\documentclass[12pt]{article}
\usepackage{amssymb,graphicx,graphics,epsfig}
\usepackage{lscape,graphics,amsmath}
\usepackage{latexsym,amsfonts}
\usepackage{axodraw}
\usepackage[english]{babel}
\usepackage{color}
\usepackage{cite}

\newcounter{diag1}
\newcounter{diag2_NC}
\newcounter{diag2_CC}

\begin{document}

\begin{center}
{\Large \bf Quantum field-theoretical description of neutrino
oscillations in magnetic field}\\
\vspace{4mm} Vadim Egorov, Igor Volobuev\\
\vspace{4mm} Skobeltsyn Institute of Nuclear Physics, Moscow State
University
\\ 119991 Moscow, Russia\\
\end{center}
\begin{abstract}
It is shown that the processes of neutrino oscillations in a
magnetic field can be consistently described in the framework of a
new quantum field-theoretical approach without
use of the neutrino flavor states. It is based on the Feynman
diagram technique with a modified distance-dependent propagator,
which takes into account the geometry of neutrino oscillation
experiments.  Processes of neutrino oscillations in a magnetic
field, where the neutrinos are detected through the weak charged-
and neutral-current interaction, have been studied and numerical
calculations  have been carried out for some specific examples.
Implications for the solar neutrinos are briefly discussed,
and formulas for the asymptotic values of the
normalized probability of solar neutrino oscillation processes are
derived, which coincide with  the observable ratio of the
measurable neutrino flux to that predicted by the standard solar
model.
\end{abstract}

\section{Introduction}
The Standard Model (SM) allows one to describe, with  high
accuracy, a great amount  of various elementary particle
interaction processes in the framework of the S-matrix formalism
and Feynman diagram technique. However, there is a belief that it
cannot describe the phenomena of neutral kaon and neutrino
oscillations, the latter being  under intense theoretical and
experimental investigation nowadays. The standard way to describe
the phenomenon is the quantum-mechanical approach in terms of
plane waves
\cite{Pontecorvo:1957cp,Gribov:1968kq,Giunti:2007ry,Bilenky:2010zza}.
Although being straightforward, it is believed to be inconsistent,
because the production of the neutrino flavor states is described
within the SM (which is a gauge field theory), whereas their
evolution is described within quantum mechanics. Such a
description seems to be eclectic, since quantum field theory
includes quantum mechanics as an indispensable part and must be
able to describe all quantum phenomena. Moreover, the production
of states without definite mass leads to violation of
energy-momentum conservation
\cite{Giunti:1993se,Grimus:1996av,Beuthe:2001rc,Cohen:2008qb}.
This problem is supposedly  solved in the framework of the
quantum-mechanical approach it terms of wave packets
\cite{Giunti:2007ry,Kayser}, although this description turns out
to be very bulky. Thus, the construction of a consistent and
convenient description of neutrino oscillations within quantum
field theory is of current interest.

The first attempt to describe neutrino oscillations in the
framework  of quantum field theory was made back in 1982  in paper
\cite{Okun1982}. In this paper, within the standard perturbative
S-matrix formalism, it was assumed that virtual neutrino mass
eigenstates were produced and detected in the charged-current
interactions with nuclei. The matrix elements of the charged weak
hadron current between the initial and final states of the nuclei
were approximated by delta functions of their positions separated
by a fixed distance, whereas all the incoming and outgoing leptons
were described by plane waves. This approximation fixed the
distance between the production and detection points of the
neutrinos and, in the momentum representation,  resulted in the
appearance of distance-dependent neutrino propagators. Hence,
neutrino oscillations were  regarded as interference  of the
amplitudes of processes mediated by different neutrino mass
eigenstates.  In the subsequent studies the delta functions, which
seemed to be a rather too rough approximation and formally
contradicted the S-matrix formalism, were replaced by localized
wave packets  for describing the states of the nuclei
\cite{Giunti:1993se,Grimus:1996av,Beuthe:2001rc,Grimus:2019hlq}.
However, the calculations in the framework of this approach turned
out to be very bulky and complicated. The reason is that the
standard perturbative formalism of S-matrix is not suitable for
describing processes taking place at finite space and time
intervals.

It is a common knowledge that the presence of external fields and
matter affects neutrino oscillations \cite{Giunti:2007ry}.  The
influence of magnetic fields on neutrino oscillations, as well as
its possible implications for the solar neutrino problem were
repeatedly considered in the framework of the standard
quantum-mechanical description based on the use of the neutrino
flavor states in papers
\cite{Cisneros:1971,Fujikawa:1980yx,Schechter:1981hw,Okun1986,Akhmedov,Popov:2019nkr,Chukhnova:2019oum}.
However, up to now, there were no papers dealing with the
description of neutrino oscillations in external fields and matter
within quantum field theory without use of the flavor states. In
the present paper, we develop such a quantum field-theoretical
description of neutrino oscillations in a magnetic field. Our
approach makes use of a modified perturbative formalism adapted
for describing processes passing at finite space and time
intervals
\cite{Volobuev:2017izt,Egorov:2017qgk,Egorov:2017vdp,Volobuev:2019zan,Egorov:2019vqv}.
The formalism is based on the Feynman diagram technique in the
coordinate representation supplemented with modified rules of
passing to the momentum representation. The latter reflect the
geometry of neutrino oscillation experiments and lead to the
Feynman propagator of virtual neutrino mass eigenstates in the
momentum representation being modified. Namely, a
distance-dependent propagator of neutrino mass eigenstates in the
momentum representation arises, while the rest of the Feynman
rules in this representation are kept intact. The description in
terms of plane waves allows one to avoid cumbersome calculations,
while catching the essence of the phenomenon.

In describing neutrino oscillations in a magnetic
field, we assume that neutrinos are produced and detected through
the charged- and neutral-current interaction with nuclei and
electrons in the absence of field, but the propagation of the
particles occurs in a region of magnetic field. In Section 2 we
give a brief review of the main ideas of the approach. In Section
3 we apply the formalism to the case of neutrino oscillation in a
magnetic field, deriving the oscillation probability. In Section 4
we consider specific examples with solar neutrinos, namely
neutrino production in $^{15}\rm O$ decay or electron capture by
${^{7} {\rm Be}}$ and detection by Ga-Ge or Cherenkov detectors,
and derive useful formulas for the asymptotic values of the
normalized probability of solar neutrino oscillation processes,
which coincide with  the observable ratio of the measurable
neutrino flux to that predicted by the standard solar model.

\section{Basics of the approach}

First, we note that neutrino oscillation experiments are
characterized by a specific geometry, where the distance between a
neutrino source and a detector is much larger than their sizes.
For this reason one can consider the source and  detector to be
pointlike and describe the oscillation process by just one
parameter of geometric origin: the distance between the centers of
the source and detector. Moreover, one can use the one-dimensional
approximation, where the neutrino momenta are directed along the
line connecting the centers, which is usually called the plane
wave approximation in the standard quantum-mechanical approach
\cite{Giunti:2007ry}.

In what follows we  use the one-dimensional approximation. Also,
we work in the framework of the minimal extension of the Standard
Model by the right neutrino singlets. The interaction Lagrangian
of the leptons takes the form
\begin{equation} \label{L_intl}
\begin{split}
L_{\rm int}^{\rm lep} =& - \frac{g }{2\sqrt{2}}\left(\sum_{i,k = 1}^3
\bar l_i \, \gamma^\mu \left(1 - \gamma^5 \right) U_{ik} \, \nu_k
\, W^{-}_\mu + {\rm h.c.}\right) +   \frac{g \sin^2\theta_{\rm w}}{
\cos\theta_{\rm w}}\sum_{i=1}^3 \bar l_i \, \gamma^\mu \,
 l_i \, Z_\mu - \\
& - \frac{g }{4 \cos\theta_{\rm w}}\sum_{i=1}^3 \bar l_i \, \gamma^\mu
\left(1 - \gamma^5 \right) l_i \, Z_\mu +   \frac{g }{4
\cos\theta_{\rm w}}\sum_{k=1}^3 \bar \nu_k \, \gamma^\mu \left(1 -
\gamma^5 \right) \nu_k \, Z_\mu \, ,
\end{split}
\end{equation}
where $l_i$ denotes the field of the charged lepton of the $i$th
generation,  $U_{ik}$ is the PMNS-matrix, $\nu_k$ stands for the
field of the neutrino state with definite mass $m_k$, and
$\theta_{\rm w}$ is the Weinberg angle.

Let us first recall the basics of the approach and consider the
case of neutrino oscillations in vacuum. To this end, we consider
a process, where a neutrino is produced and detected through the
charged-current interaction with nuclei. In the lowest order of
perturbation theory the process is described by the following
diagram: \vspace{0.3cm}
\begin{figure}[h]
\begin{center}
\ \ \ \includegraphics[width=0.3965\linewidth]{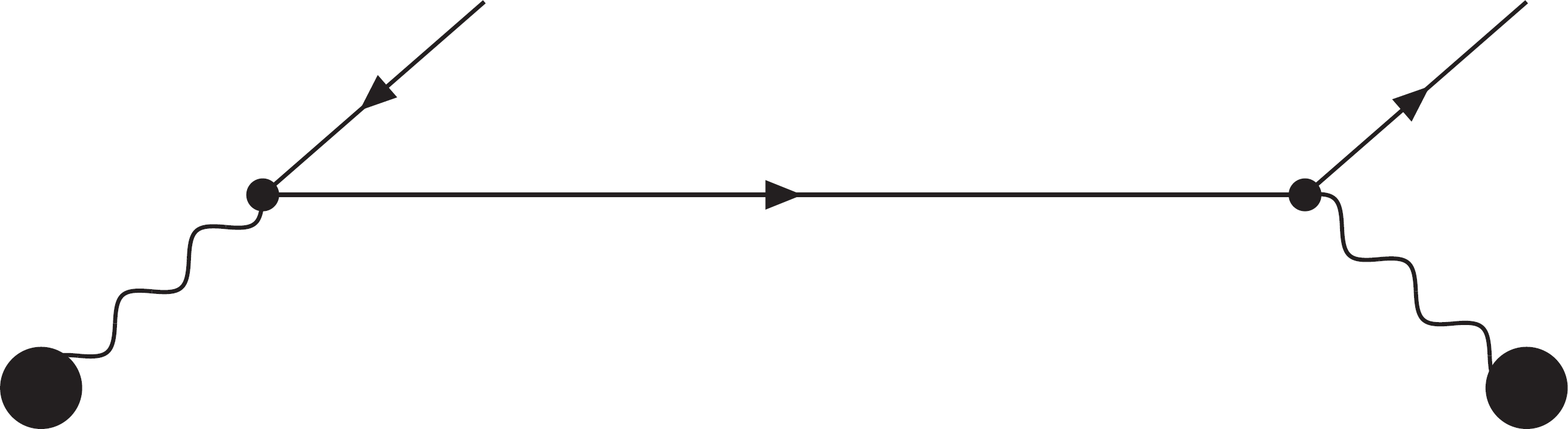}
\end{center}
\end{figure}
\vspace*{-3.1cm}
\begin{center}
\begin{picture}(197,87)(0,0)
\Text(70.0,94.0)[l]{$e^+ ( q )$}\ArrowLine(67.5,88.0)(40.5,64.5)
\Text(33.5,65.5)[r]{$x$} \Photon(13.5,41.0)(40.5,64.5){2}{3.0}
\Text(53.5,48.5)[r]{$W^+$} \Vertex (13.5,41.0){5} \Vertex
(40.5,64.5){2} \ArrowLine(40.5,64.5)(167.5,64.5) \Vertex
(167.5,64.5){2} \Text(104.8,70.5)[b]{$\nu_i ( p_{\rm n} )$}
\ArrowLine(167.5,64.5)(194.5,88.0) \Text(197.5,94.0)[l]{$e^- ( k
)$} \Text(175.0,64.5)[l]{$y$}
\Photon(167.5,64.5)(194.5,41.0){2}{3.0} \Vertex (194.5,41.0){5}
\Text(177.0,48.5)[r]{$W^+$}
\Text(332.8,60.5)[b]{\addtocounter{equation}{1}(\arabic{equation})}
\setcounter{diag1}{\value{equation}}
\label{diag1}
\end{picture}
\end{center}
\vspace{-1.349cm}
The points of production $x$ and  detection $y$
are supposed to be separated by a fixed macroscopic distance $L$
along a unit vector $\vec n, \, \vec y - \vec x =  L \vec n$.
The intermediate virtual neutrino mass eigenstate
is described by the Feynman propagator in the coordinate
representation. The amplitude corresponding to the diagram must
be summed over all the three neutrino mass eigenstates, $i =
1,2,3$.

The initial and final nuclei and particles are
assumed to be  described by plane waves. Their 4-momenta are
denoted as it is shown in the diagram: $q$, $k$, $p_{\rm n}$
correspond to the positron, electron and intermediate virtual
neutrino, respectively. The filled circles in the diagram
represent the matrix elements of the weak charged hadron current
between the states of initial nuclei 1 $\left(^{A_1}_{Z_1} {\rm
X}\right) $  and 2 $\left(^{A_2}_{Z_2} {\rm X}\right) $  and final
nuclei $1^\prime$ $\left(^{A_1}_{Z_1 - 1} {\rm X}\right)$ and
$2^\prime$ $\left(^{A_2}_{Z_2 + 1} {\rm X}\right)$:
\begin{equation} \label{hadron_currents}
\begin{split}
j_\mu^{(1)} \left( {\vec P^{(1)}, \vec P^{(1^\prime)} } \right) &= \left <^{A_1}_{Z_1 - 1} {\rm X} \left( {\vec P^{(1^\prime)} } \right) \right| j_\mu^{({\rm h})} \left| ^{A_1}_{Z_1} {\rm X} \left( {\vec P^{(1)}} \right) \right>, \\
j_\rho^{(2)} \left( {\vec P^{(2)}, \vec P^{(2^\prime)} } \right)
&= \left <^{A_2}_{Z_2 + 1} {\rm X} \left( {\vec P^{(2^\prime)} }
\right) \right| j_\rho^{({\rm h})} \left| ^{A_2}_{Z_2} {\rm X}
\left( {\vec P^{(2)}} \right) \right>,
\end{split}
\end{equation}
the nuclei 4-momenta being denoted by $P^{(l)} = \left( {E^{(l)},
\vec P^{(l)} } \right)$, $l= 1,1^\prime,2,2^\prime$.

The amplitude of the process in the coordinate representation can
be  constructed using the Feynman rules formulated, for example,
in textbook \cite{BOSH}. The passing to the
momentum representation is performed by integrating the amplitude
with respect to $x$ and $y$ over Minkowski space. However, such
a straightforward integration would result in losing the
information about the space-time interval between the production
and detection events. Thus, in order to be able to describe
processes passing at finite space and time interval one has to fix
somehow the distance between the neutrino production and detection
points. To this end, in paper \cite{Okun1982} the matrix elements
of hadron currents (\ref{hadron_currents}) in the coordinate
representation were assumed to be proportional to  the delta
functions $\delta(\vec x - \vec x_1), \,\, \delta(\vec y - \vec
x_2) $ with $\vec x_2 - \vec x_1$ fixed, $\vec x_2 - \vec x_1 = L
\vec n  ,$ which, in the approximation of Fermi's interaction,
gave the desired result.

In our approach, instead of fixing the positions of the initial
and final nuclei, we just fix the distance $L$ between the
interaction points along the unit vector $\vec n$ directed from
the source to the detector by introducing the delta function
$\delta (\vec n (\vec y -\vec x) - L )$ into the integrand, which
gives a generalization of the standard perturbative formalism to
the case of processes passing at finite distances.

The introduction of the delta function is formally equivalent to
replacing the standard Feynman propagator $S^{\rm c}_i(y - x)$ of
the neutrino mass eigenstate $\nu_i$ in the coordinate
representation by $S^{\rm c}_i(y - x) \, \delta (\vec n (\vec y
-\vec x) - L )$. The Fourier transform of this expression was
called in paper \cite{Volobuev:2017izt} \textit{the
distance-dependent propagator of the neutrino mass eigenstate
$\nu_i$ in the momentum representation}:
\begin{equation} \label{dist_dep_prop}
  S^{\rm c}_i \left( {p, \vec n, L} \right) \equiv \int {{\rm d}^4 z \, {\rm e}^{{\rm i}pz} \,
  S^{\rm c}_i \left( z \right) \delta \left( {\vec n \vec z - L} \right)} \,  .
\end{equation}
This integral can be evaluated exactly:
\begin{equation}\label{prop_L_mom_4a}
S^{\rm c}_i (p, \vec n, L) =  {\rm i}\, \frac{\hat p + \vec \gamma \vec
 n \left(\vec p \vec n - \sqrt{(\vec p \vec n)^2 + p^2 - m^2_i}\right) + m_i }
{2 \sqrt{(\vec p \vec n)^{\,2} + p^2 - m^2_i + {\rm i} \varepsilon}}\, {\rm e}^{-{\rm i}\left(\vec p
\vec n - \sqrt{(\vec p \vec n)^{\,2} + p^2 - m^2_i}\,\right) L} \,,
\end{equation}
where $\hat p = \gamma^\mu p_\mu$.

According to the Grimus-Stockinger theorem \cite{Grimus:1996av}
the virtual particles propagating over macroscopic distances are
almost on the mass shell, and for the momenta $\vec p$ satisfying
\\ $\left| {p^2 - m_i^2 } \right| / (\vec p \vec n )^{\, 2} \ll 1$
the distance-dependent propagator can be brought to the simple
form
\begin{equation} \label{dist_dep_prop_on-shell}
S_i^{\rm c} \left( {p, \vec n, L} \right) = {\rm i}\, \frac{{\hat p +
m_i }}{{2 {\vec p \vec n } }}\,{\rm e}^{{\rm i}\frac{{p^2 - m_i^2
}}{{2 {\vec p \vec n }}}L}\, .
\end{equation}
In particular, this approximation is always valid for the neutrino
momenta $\vec p$ directed along the vector $\vec n$, $\vec p\vec n
= |\vec p|,$  which are the only momenta needed for calculating
the amplitudes.

Distance-dependent propagator (\ref{dist_dep_prop_on-shell})
differs from the one found in paper \cite{Okun1982}: there the
propagator is the projection operator ${\hat p + m_i }$ multiplied
by a spherical wave, whereas in our case this operator is
multiplied by the exponential similar to that appearing in the
standard plane wave approximation. It is also necessary to note
that a distance-dependent propagator very similar to ours has been
obtained in recent paper \cite{Fujikawa:2020mei} within a path
integral approach.

Finally, using expression (\ref{dist_dep_prop_on-shell}) instead
of the usual Feynman propagator for constructing the amplitude in
the momentum representation  allows one to consistently describe
neutrino oscillations in vacuum
\cite{Egorov:2017qgk,Egorov:2017vdp,Volobuev:2019zan,Egorov:2019vqv}.
It is worth noting that in this case one can also use the
time-dependent propagator, which is obtained by fixing the time
interval $T$ between the neutrino production and detection events.
This is due to the fact that, for the neutrinos almost on the mass
shell, the space and time intervals  are related by the standard
formula
\begin{equation}
L = \frac{|\vec p|}{p^0}T.
\end{equation}
However, this relation is no longer true for neutrinos in an
external field, and one has to use  distance-dependent propagator
(\ref{dist_dep_prop_on-shell}) in this case.

\section{Neutrino oscillations in a magnetic field}

Now let us turn a background electromagnetic field on. Neutrinos
are  able to interact with it through quantum loops, which
provides the neutrino with, among others, anomalous dipole
magnetic moment. We will take it into account but neglect the
transition moments, which are usually assumed to be much smaller.
This means that the equations of motion for different neutrino
mass eigenstates are not coupled. Thus, the equation of motion of
a neutrino mass eigenstate in an external electromagnetic field
takes the form
\begin{equation} \label{eq_of_mot_F}
    \left( {{\rm i}\gamma ^\mu  \partial _\mu   - m_i  - \frac{{\rm i}}{2}\mu _0 m_i F_{\mu \nu } \gamma ^{\mu \nu } } \right)\nu _i \left( x \right) = 0 \, ,
\end{equation}
where the magnetic moment of the $i$th neutrino mass eigenstate,
proportional to its mass, is $\mu_i = \mu_0 m_i$
and $\gamma ^{\mu \nu }  = \frac{1}{2} \left[ {\gamma ^\mu ,\gamma
^\nu  } \right]$.  In the Standard Model, the parameter $\mu _0 =
{{3eG_{\rm F} } \mathord{\left/ {\vphantom {{3eG_{\rm F} } {8\sqrt
2 \pi ^2 }}} \right. \kern-\nulldelimiterspace} {8\sqrt 2 \pi ^2
}}$, which gives, for the experimentally allowed neutrino masses,
the neutrino magnetic moments at least 10 orders of magnitude
smaller than the Bohr magneton, but they may be much larger in  SM
extensions.

For a homogeneous electromagnetic field, Green's function of
equation (\ref{eq_of_mot_F}) in the momentum representation reads
\begin{equation} \label{Green}
\begin{split}
 S_i^{\rm c} \left( p \right) =& \ {\rm i}\left\{ {\left( {p^2  - m_i^2 } \right)\left( {p^2  - m_i^2  + {\rm i}\varepsilon } \right) - \mu _0^2 m_i^2 \left[ {\left( {p^2  + m_i^2 } \right)F_{\mu \nu } F^{\mu \nu }  - 4F_{\mu \nu } p^\nu  F^{\mu \sigma } p_\sigma  } \right] +
\vphantom{\frac{1}{4}} } \right. \\
& + \frac{1}{4}\mu _0^4 m_i^4 \left[ {\left( {F_{\mu \nu } F^{\mu
\nu } } \right)^2  + \left( {F_{\mu \nu } \tilde F^{\mu \nu } }
\right)^2 } \right] \bigg\}^{ - 1}   \left\{ {\left( {p^2  - m_i^2
} \right)\left( {\hat p + m_i } \right) -
\vphantom{\frac{1}{4}} } \right. \\
& - \frac{1}{2}\mu _0^2 m_i^2 F_{\mu \nu } F^{\mu \nu } \left( {\hat p - m_i } \right) - 2\mu _0^2 m_i^2 F_{\mu \nu } F^{\nu \sigma } p_\sigma  \gamma ^\mu   + 2\mu _0 m_i^2 \tilde F_{\mu \nu } p^\nu  \gamma ^\mu  \gamma ^5  + \\
& + {\rm i}\mu _0 m_i \left[ {\frac{1}{2}\left( {p^2  + m_i^2 } \right)F_{\mu \nu }  - \frac{1}{4}\mu _0^2 m_i^2 F^{\rho \sigma } \left( {F_{\rho \sigma } F_{\mu \nu }  + \tilde F_{\rho \sigma } \tilde F_{\mu \nu } } \right) - 2F_{\mu \rho } p^\rho  p_\nu  } \right]\gamma ^{\mu \nu } - \\
& - \frac{{\rm i}}{2}\mu _0^2 m_i^3 F_{\mu \nu } \tilde F^{\mu \nu
} \gamma ^5 \bigg\} \, .
\end{split}
\end{equation}
Here $\tilde F^{\mu \nu }  =  -
\frac{1}{2}\varepsilon ^{\mu \nu \rho \sigma } F_{\rho \sigma }$, $\varepsilon ^{0 1 2 3} = -1$.

First, we consider a homogeneous magnetic field $\vec H$: $F_{\mu
\nu }  = \varepsilon _{\mu \nu k0} H^k$, $k = 1,2,3$. Then the
neutrino dispersion relation following from the denominator of
Green's function (\ref{Green}),
\begin{equation} \label{disp_rel}
\left( {p^0 } \right)^2  = \vec p^{\, 2}  + m_i^2  + \mu _0^2 m_i^2 \vec H^2  \pm 2\mu _0 m_i \sqrt {\vec p^{\, 2} \vec H_ \bot ^2  + m_i^2 \vec H^2 } \,,
\end{equation}
where $\vec H_ \bot$ denotes the component of the
magnetic field $\vec H$ transverse to the direction of neutrino
propagation $\vec n = {{\vec p} \mathord{\left/ {\vphantom {{\vec
p} {\left| {\vec p} \right|}}} \right. \kern-\nulldelimiterspace}
{\left| {\vec p} \right|}}$, coincides, mutatis mutandis, with the
dispersion relation for the neutron in a magnetic field, which was
first derived in paper \cite{Ternov1965} and recently reproduced
within the standard approach in paper \cite{Popov:2019nkr} for the
neutrinos.

Since the neutrino magnetic moment is extremely small,  $\mu _0^2
m_i^2 \vec H^2 \ll \vec p ^{\, 2}$, we neglect the terms of order
2 and higher in $\mu_0$. Substituting the Green's function in the
coordinate representation in  definition (\ref{dist_dep_prop}) of
the distance-dependent propagator, taking the neutrino momentum to
be parallel to $\vec n$ and neglecting also the neutrino masses
everywhere except in the exponential, we arrive at \textit{the
distance-dependent propagator of the neutrino mass eigenstate in a
homogeneous magnetic field in the momentum representation}:
\begin{equation} \label{dist_dep_prop_field}
S_i^{\rm c} \left( {p, L,\vec H} \right) = {\rm i}\, \frac{{\hat
p\left( {1 - {\rm i}\vec \gamma \vec j} \right)}}{{4\left| {\vec
p} \right| }}\,{\rm e}^{{\rm i}\frac{{p^2  - m_i^2 + 2 \mu _0 m_i
\left| {\vec p} \right| H_ \bot   }}{{2\left| {\vec p} \right|}}
L}  + {\rm i}\,\frac{{\hat p\left( {1 + {\rm i}\vec \gamma \vec j}
\right)}}{{4\left| {\vec p} \right| }}\,{\rm e}^{{\rm i}\frac{{p^2
- m_i^2 - 2 \mu _0 m_i \left| {\vec p} \right| H_ \bot
}}{{2\left| {\vec p} \right|}} L} .
\end{equation}
Here $H_ \bot = | { \vec H _\bot } |$, we suppose $m_i^2 \vec H^2
\ll \vec p^{\, 2} \vec H_ \bot ^2$, and
\begin{equation}
\vec j \equiv \frac{{\left[ {\vec n \times \vec h} \right]}}{{\sqrt {1 - \left( {\vec n\vec h} \right)^2 } }}, \qquad \vec h \equiv \frac{{\vec H}}{{\left| {\vec H} \right|}}, \qquad \vec j^2  = 1,
\end{equation}
the square brackets denoting the vector product. Comparing
expressions (\ref{dist_dep_prop_field}) and
(\ref{dist_dep_prop_on-shell}) one can conclude that,  in the
magnetic field,  each neutrino mass eigenstate splits into two
states corresponding to two possible spin orientations and
energies. This effect is in full agreement with the
quantum-mechanical expectations. The numerators in the exponents,
 $p^2 - m_i^2 \mp 2 \mu _0 m_i \left| {\vec p} \right|
H_ \bot $, measure the deviation of the virtual neutrinos from the
mass shell, which is consistent with  dispersion relation
(\ref{disp_rel}) in our approximation.

We see that the distance-dependent propagator essentially depends
only  on the transverse component of a magnetic field. In what
follows, we will neglect the longitudinal component of the field
and assume the magnetic field to be transverse. Further,
propagator (\ref{dist_dep_prop_field}) has been derived for the
case of a constant homogeneous magnetic field. Nevertheless, it
can be used for the transverse magnetic field, the magnitude of
which varies along the neutrino path adiabatically, i.e., if the
condition
\begin{equation}\label{adiabat}
\left| {\mu_0 m_{\max} (\vec n \vec \nabla) \left|{ \vec H } \right|
} \right| \ll \frac{|\vec p|}{d}
\end{equation}
is fulfilled, where $d$ is the characteristic size of the field
region and $m_{\max}$ is the largest neutrino mass.
This condition refines the adiabaticity
condition for the magnetic field introduced earlier in paper
\cite{Okun1986}. It takes into account the neutrino magnetic
moments, the size of the field region and guarantees that the term
$({p^2 - m_i^2})/2 | {\vec p} |$ in the exponential can
be considered to be constant along the path. However, since now
the magnetic field varies along the path, the field $H_\bot $ in
formula (\ref{dist_dep_prop_field}) should be replaced by the
mean field
\begin{equation}\label{mean_field}
\overline H = \frac{1}{L} \int\limits_0^L H\left( {l} \right) {\rm d} l.
\end{equation}

Now we will use distance-dependent propagator
(\ref{dist_dep_prop_field}) with these amendments to calculate the
probability of an actual process. Let us first consider a process
analogous to the one described by diagram (\arabic{diag1}), where
the neutrinos are produced and detected in the absence of field,
but propagate through a region of  magnetic field (for which the
assumptions above are true). In what follows we will work in the
approximation of Fermi's interaction. Then the amplitude of the
process in the momentum representation, where the distance is
fixed and equals $L$, looks as follows:
\begin{equation} \label{amplitude}
    \begin{split}
M = \ & {- {\rm i}}\frac{{G_{\rm F}^2 }}{{8 \left| { \vec p_{\rm n} } \right| }} \, j_\rho ^{(2)} \left( {\vec P^{(2)} ,\vec P^{(2')} } \right)\bar u\left( \vec k \right)\gamma ^\rho  \left( {1 - \gamma ^5 } \right)\hat p_{\rm n}  \times  \\
&\times \sum\limits_{i = 1}^3 {\left| {U_{1i} } \right|^2 \left[ {\left( {1 - {\rm i}\vec \gamma \vec j} \right){\rm e}^{{\rm i}\frac{{p_{\rm n}^2  - m_i^2  + 2\mu _0 m_i \left| {\vec p_{\rm n} } \right|\overline H  }}{{2\left| {\vec p_{\rm n} } \right|}}L}  + \left( {1 + {\rm i}\vec \gamma \vec j} \right){\rm e}^{{\rm i}\frac{{p_{\rm n}^2  - m_i^2  - 2\mu _0 m_i \left| {\vec p_{\rm n} } \right|\overline H  }}{{2\left| {\vec p_{\rm n} } \right|}}L} } \right]}  \times  \\
&\times \gamma ^\mu  \left( {1 - \gamma ^5 } \right)v\left( \vec q
\right)j_\mu ^{(1)} \left( {\vec P^{(1)} ,\vec P^{(1')} } \right).
\end{split}
\end{equation}
Here we omit fermion polarization indices for  brevity; the
notations for the momenta are the same as in diagram
(\arabic{diag1}) explained in the paragraph above formula
(\ref{hadron_currents}). Again, neutrino masses are neglected
everywhere, except in the exponential.

The squared modulus of the amplitude (\ref{amplitude}), averaged
over  the polarizations of the incoming nuclei and summed over the
polarizations of the outgoing particles and nuclei (the operation
of averaging and summation is denoted by the angle brackets),
factorizes in the approximation of zero neutrino masses. The
latter means that we also neglect the terms proportional to
$p_{\rm n}^2$, which, due to the Grimus-Stockinger theorem
\cite{Grimus:1996av}, is of the order of neutrino masses squared.
In (\ref{amplitude}), the terms containing the vector $\vec j$
vanish, and we arrive at the result
\begin{equation} \label{sqr_amp}
\left\langle {\left| M \right|^2 } \right\rangle  = \left\langle
{\left| {M_{\rm P} } \right|^2 } \right\rangle \left\langle
{\left| {M_{\rm D} } \right|^2 } \right\rangle \frac{1}{{4\vec
p_{\rm n}^{\, 2} }} \, P_{ee} \left( {\left| {\vec p_{\rm n} }
\right|,L,\overline H} \right),
\end{equation}
where
\begin{equation}
\left\langle {\left| M_{\rm P} \right|^2 } \right\rangle = 4G_{\rm F}^2 \left[ -{g^{\mu \nu } \left( {q p_{\rm n} } \right)  + q^\mu  p_{\rm n}^\nu   + p_{\rm n} ^\mu  q^\nu   - {\rm i} \varepsilon ^{\mu \nu \rho \sigma } q_\rho \left( { p_{\rm n} } \right) _\sigma } \vphantom{\left( { p_{\rm n} } \right) _\rho} \right] W_{\mu \nu }^{(1)}
\end{equation}
is the squared modulus of the amplitude of the production process;
\begin{equation}
\left\langle {\left| M_{\rm D} \right|^2 } \right\rangle = 4G_{\rm F}^2 \left[- {g^{\mu \nu } \left( {p_{\rm n} k} \right) + p_{\rm n}^\mu  k^\nu   + k^\mu  p_{\rm n}^\nu  - {\rm i} \varepsilon ^{\mu \nu \rho \sigma } \left( { p_{\rm n} } \right) _\rho  k_\sigma  } \right] W_{\mu \nu }^{(2)}
\end{equation}
is the squared modulus of the amplitude of the detection process;
\begin{equation}
 W_{\mu \nu }^{(l)} = W_{\mu \nu }^{(l,{\rm S})} + {\rm i} W_{\mu \nu }^{(l,{\rm A})} = \left\langle {j_\mu ^{(l)} \left( {j_\nu ^{(l)} } \right)^ +  } \right\rangle , \quad l=1,2,
\end{equation}
are the nuclear tensors characterizing the interaction of nuclei
1, $1^\prime$ and 2, $2^\prime$ with the leptons, their symmetric
parts $W_{\mu \nu }^{(l,{\rm S})}$ being real and the
anti-symmetric ones ${\rm i} W_{\mu \nu }^{(l,{\rm A})}$ being
imaginary;
\begin{equation} \label{prob_osc}
    \begin{split}
     P_{ee} \left( {\left| {\vec p_{\rm n} } \right|,L, \overline H} \right)&  = \ 1 - \sum\limits_{\scriptstyle i,k = 1 \hfill \atop
        \scriptstyle k < i \hfill}^3 {\left| {U_{1i} } \right|^2 \left| {U_{1k} } \right|^2 \left\{ {\sin ^2 \left[ {\left( {\frac{{\Delta m_{ik}^2 }}{{4\left| {\vec p_{\rm n} } \right|}} - \frac{{\mu _0 \Delta m_{ik} \overline H  }}{2}} \right)L} \right] + } \right.}  \\
        &+ \sin ^2 \left[ {\left( {\frac{{\Delta m_{ik}^2 }}{{4\left| {\vec p_{\rm n} } \right|}} - \frac{{\mu _0 \Sigma m_{ik} \overline H  }}{2}} \right)L} \right] + \sin ^2 \left[ {\left( {\frac{{\Delta m_{ik}^2 }}{{4\left| {\vec p_{\rm n} } \right|}} + \frac{{\mu _0 \Sigma m_{ik} \overline H  }}{2}} \right)L} \right] +  \\
        & \left. { + \sin ^2 \left[ {\left( {\frac{{\Delta m_{ik}^2 }}{{4\left| {\vec p_{\rm n} } \right|}} + \frac{{\mu _0 \Delta m_{ik} \overline H  }}{2}} \right)L} \right]} \right\} - \sum\limits_{i = 1}^3 {\left| {U_{1i} } \right|^4 \sin ^2 \left( {\mu _0 m_i \overline H  L} \right)}
    \end{split}
\end{equation}
is the neutrino oscillation probability depending on the neutrino
momentum $ {\vec p_{\rm n} }$, $L$ is the distance between the
source and detector and $\overline H$ denotes the mean value of
the transverse magnetic field; we also introduce the notations
similar to the usual ones $\Delta m_{ik}^2 \equiv m_i^2 - m_k^2$:
\begin{equation}
\Delta m_{ik} \equiv m_i  - m_k , \qquad \Sigma m_{ik} \equiv m_i  + m_k .
\end{equation}
In the case of two neutrino flavors formula (\ref{prob_osc}) is
consistent with the ones derived in papers
\cite{Popov:2019nkr,Chukhnova:2019oum}  for a homogeneous magnetic
field, although it looks different.

We note that, since the neutrinos are produced and detected in the
absence of magnetic field and the field is transverse to the
neutrino path, the energy and momentum are conserved in the
process under consideration. Following the prescription formulated
in papers \cite{Egorov:2017qgk,Egorov:2017vdp,Volobuev:2019zan},
in order to find the probability of the process, before
integrating over the phase volume of the final particles, one must
multiply expression (\ref{sqr_amp}) not only by the delta function
of energy-momentum conservation, but also by a delta function,
which would guarantee that the momentum  $ \vec p_{\rm n}$ of the
intermediate neutrinos is directed along the vector $\vec n$.
Since we have calculated the squared modulus of the amplitude in
the approximation of massless neutrinos, it is natural to
calculate the probability in the same approximation, i.e., to
choose $ p_{\rm n}^2 = 0$.  A special value of $p_{\rm n}$
satisfying these conditions  will be denoted by $p$, where $\vec
p$ is directed from the source to the detector and $p^2 = 0$.

Thus, we  multiply the  squared modulus of the amplitude by the
delta function of energy-momentum conservation $\left( {2\pi }
\right)^4 \delta \big( P^{(1)}  + P^{(2)}  - P^{(1')}  - q -
P^{(2')}  - k \big)$  and also by the delta function $2\pi \,
\delta \left( {P^{(1)} - P^{(1')} - q - p} \right)$, as well as
substitute $p$ instead of $p_{\rm n}$ everywhere in
(\ref{sqr_amp}) \cite{Egorov:2017qgk}. In so doing, we fix the
momentum of the intermediate neutrinos, whose direction is
determined by the source-detector relative position, and, after
the integration over the phase volume, find the differential
probability ${{{{\rm d}^3 W} \mathord{\left/ {\vphantom {{{\rm
d}^3 W} {{\rm d}^3 p}}} \right. \kern-\nulldelimiterspace} {{\rm
d}^3 p}}}$ of the process with a definite neutrino momentum. Since
the experimental situation determines only the direction of
neutrino momentum, but not its magnitude, we must also integrate
${{{{\rm d}^3 W} \mathord{\left/ {\vphantom {{{\rm d}^3 W} {{\rm
d}^3 p}}} \right. \kern-\nulldelimiterspace} {{\rm d}^3 p}}}$ with
respect to $\left| {\vec p } \right|$ over the admissible values.
The final probability of the process under consideration reads:
\begin{equation} \label{final_prob}
    \frac{{{\rm d}W}}{{{\rm d}\Omega }} = \int\limits_{\left| {\vec p} \right|_{\min } }^{\left| {\vec p} \right|_{\max } } {\frac{{{\rm d}^3 W}}{{{\rm d}^3 p}}\left| {\vec p} \right|^2 {\rm d} | {\vec p} |}  = \int\limits_{\left| {\vec p} \right|_{\min } }^{\left| {\vec p} \right|_{\max } } {\frac{{{\rm d}^{\rm 3} W_{\rm P} }}{{{\rm d}^3 p}} \, W_{\rm D} \, P_{ee} \left( {\left| {\vec p} \right|,L,\overline H  } \right)\left| {\vec p} \right|^2 {\rm d} | {\vec p} |} .
\end{equation}
Here
\begin{equation} \label{decay_prob}
    \frac{{{\rm d}^3 W_{\rm P} }}{{{\rm d}^3 p}} = \frac{1}{{2E^{(1)} }} \frac{1}{{\left( {2\pi } \right)^3 2p^0 }} \int {\frac{{{\rm d}^3 q}} {{\left( {2\pi } \right)^3 2q^0 }} \frac{{{\rm d}^3 P^{(1')} }} {{\left( {2\pi } \right)^3 2E^{(1')} }} \left. { \left\langle {\left| {M_{\rm P} } \right|^2 } \right\rangle }\right|_{p_{\rm n}  = p}  \left( {2\pi } \right)^4 \delta \left( {P^{(1)}  - P^{(1')}  - q - p} \right)}
\end{equation}
is the differential probability of the decay of nucleus 1  into
nucleus $1^\prime$, positron and a massless fermion with momentum
$\vec p$, which coincides with the sum of differential
probabilities of the decay of nucleus 1 into nucleus $1^\prime$,
positron and all three neutrino mass eigenstates;
\begin{equation}
W_{\rm D} = \frac{1}{{2E^{(2)} 2p^0 }}\int {\frac{{{\rm d}^3 k}}{{\left( {2\pi } \right)^3 2k^0 }} \frac{{{\rm d}^3 P^{(2')} }}{{\left( {2\pi } \right)^3 2E^{(2')}}} \left. {\left\langle {\left| {M_{\rm D} } \right|^2 } \right\rangle }\right|_{p_{\rm n}  = p} \left( {2\pi } \right)^4 \delta \left( {P^{(2)}  + p - P^{(2')}  - k} \right) }
\end{equation}
is the probability of the scattering process of the massless
fermion and nucleus 2 with the production of nucleus $2^\prime$
and an electron, which coincides with the sum of the probabilities
of the scattering processes of all three neutrino mass eigenstates
and nucleus 2. The lower limit of integration $\left| {\vec p}
\right|_{\min }$ is determined by the threshold of the
registration process and the upper one $\left| {\vec p}
\right|_{\max }$ is determined by the energy-momentum conservation
in the production vertex. In what follows, we assume the initial
nuclei 1 and 2 to be at rest and put their momenta $\vec P^{(1)}$,
$\vec P^{(2)}$ equal to zero, then the integration limits are
given by \cite{BK}
\begin{equation} \label{p_min_max}
\left| {\vec p} \right|_{\min }  = \frac{{\left( {M_{2'}  + m_e} \right)^2  - M_2^2 }}{{2M_2 }}, \qquad \left| {\vec p} \right|_{\max }  = \frac{{M_1^2  - \left( {M_{1'}  + m_e} \right)^2 }}{{2M_1 }},
\end{equation}
where $M_l$, $l= 1,1^\prime,2,2^\prime$, are the nuclear masses
and $m_e$ is the electron mass. In the next section we will apply
formula (\ref{final_prob}) to specific neutrino oscillation
processes.

It is easy to verify that one arrives at an expression of the same
form as (\ref{final_prob}), if the production process is not a
nuclear decay but  electron capture:
\begin{equation}\label{nuc_e}
e^-  +  {^{A_1}_{Z_1} {\rm X}}  \to {^{A_1}_{Z_1 - 1} {\rm X}}  +
\nu_i .
\end{equation}
The only difference is that the differential decay probability
(\ref{decay_prob})  must be replaced by the differential
scattering probability. Oscillating factor (\ref{prob_osc}) does
not change. However, since the reaction has only two particles in
the final state, the magnitudes of the momenta are already fixed
by energy-momentum conservation. Hence, the differential
probability ${{{{\rm d}^3 W_{\rm P}} \mathord{\left/ {\vphantom
{{{\rm d}^3 W_{\rm P}} {{\rm d}^3 p}}} \right.
\kern-\nulldelimiterspace} {{\rm d}^3 p}}}$ for reaction
(\ref{nuc_e}) is singular, and the integration with respect to
neutrino momentum magnitude leads to
\begin{equation} \label{final_prob_2body}
\frac{{{\rm d}W}}{{{\rm d}\Omega }} = \int\limits_{\left| {\vec p}
\right|_{\min } }^{\left| {\vec p} \right|_{\max } } {\frac{{{\rm
d}^{\rm 3} W_{\rm P} }}{{{\rm d}^3 p}} \, W_{\rm D} \, P_{ee}
\left( {\left| {\vec p} \right|,L,\overline H  } \right)\left|
{\vec p} \right|^2 {\rm d} | {\vec p} |} = \frac{{{\rm d}W_{\rm P}
}}{{{\rm d}\Omega }}\left. {W_{\rm D} } \right|_{\left| {\vec p}
\right| = \left| {\vec p} \right|^ *  } P_{ee} \left( {\left|
{\vec p} \right|^ *  ,L,\overline H  } \right),
\end{equation}
where $\left| {\vec p} \right|^ *$ is the neutrino momentum
magnitude selected by the energy-momentum conservation in the
production process and ${{{{\rm d} W_{\rm P}} \mathord{\left/
{\vphantom {{{\rm d} W_{\rm P}} {{\rm d} \Omega}}} \right.
\kern-\nulldelimiterspace} {{\rm d} \Omega}}}$ is the differential
probability of production of the neutrino moving in the direction
of the detector. Since the initial particles always have a
momentum distribution, one must average probability
(\ref{final_prob_2body}) over the momenta of these particles.

In the same way we can consider a process, where the neutrino is
detected through  both the charged- and neutral-current
interactions with an electron. The process is described by the
diagrams \vspace{0.63cm}
\begin{figure}[h!!!]
\begin{center}
\hspace{-1.25cm} \includegraphics[width=0.4434\linewidth]{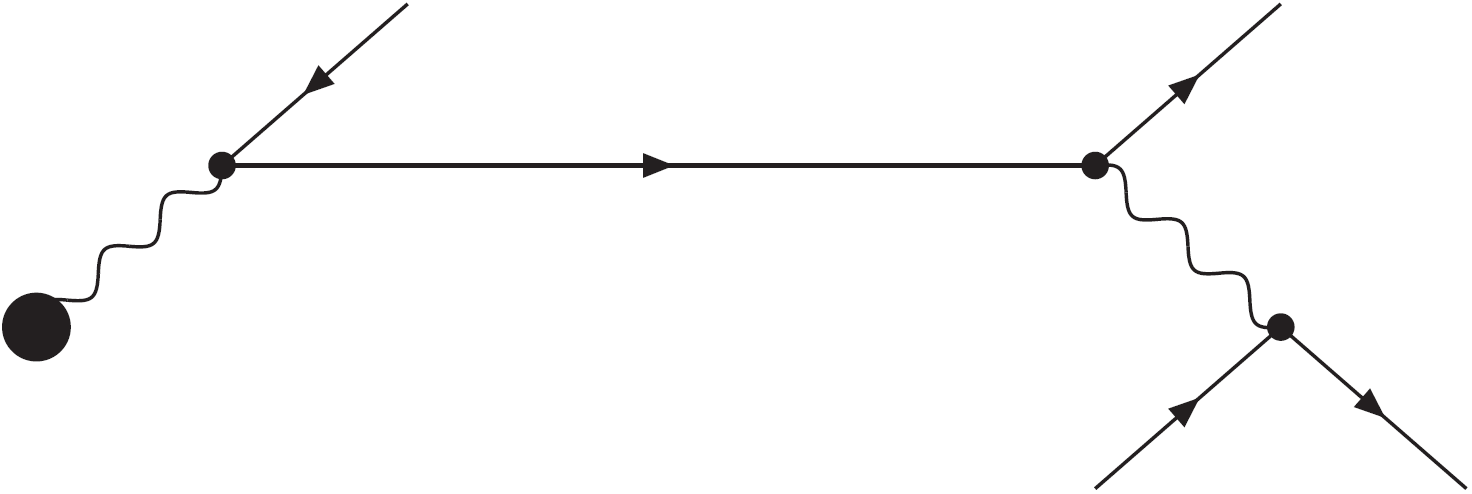}
\end{center}
\end{figure}
\vspace*{-3.6cm}
\begin{center}
\begin{picture}(193,81)(35,0)
\Text(70.0,94.0)[l]{$e^+ ( q)$}\ArrowLine(67.5,88.0)(40.5,64.5)
\Text(33.5,65.5)[r]{$x$} \Photon(13.5,41.0)(40.5,64.5){2}{3.0}
\Text(53.5,48.5)[r]{$W^+$} \Vertex (13.5,41.0){5} \Vertex
(40.5,64.5){2} \ArrowLine(40.5,64.5)(167.5,64.5) \Vertex
(167.5,64.5){2} \Text(104.8,70.5)[b]{$\nu_i ( p_{\rm n} )$}
\ArrowLine(167.5,64.5)(194.5,88.0) \Text(198.0,94.0)[l]{$\nu_i ( k_2 )$}
\Text(175.0,64.5)[l]{$y$} \Text(166.0,48.5)[l]{$Z$}
\Photon(167.5,64.5)(194.5,41.0){2}{3.0} \Vertex (194.5,41.0){2}
\ArrowLine(167.5,17.5)(194.5,41.0) \Text(132.5,17.5)[l]{$e^- ( k_1 )$}
\ArrowLine(194.5,41.0)(221.5,17.5) \Text(225.0,17.5)[l]{$e^- ( k )$}
\Text(363.0,60.5)[b]{\addtocounter{equation}{1}(\arabic{equation})}
\setcounter{diag2_NC}{\value{equation}}
\end{picture}
\end{center}
\newpage
\textcolor{white}{.}
\vspace{-1.155cm}
\begin{figure}[h!!!]
\begin{center}
\hspace{-1.25cm} \includegraphics[width=0.4434\linewidth]{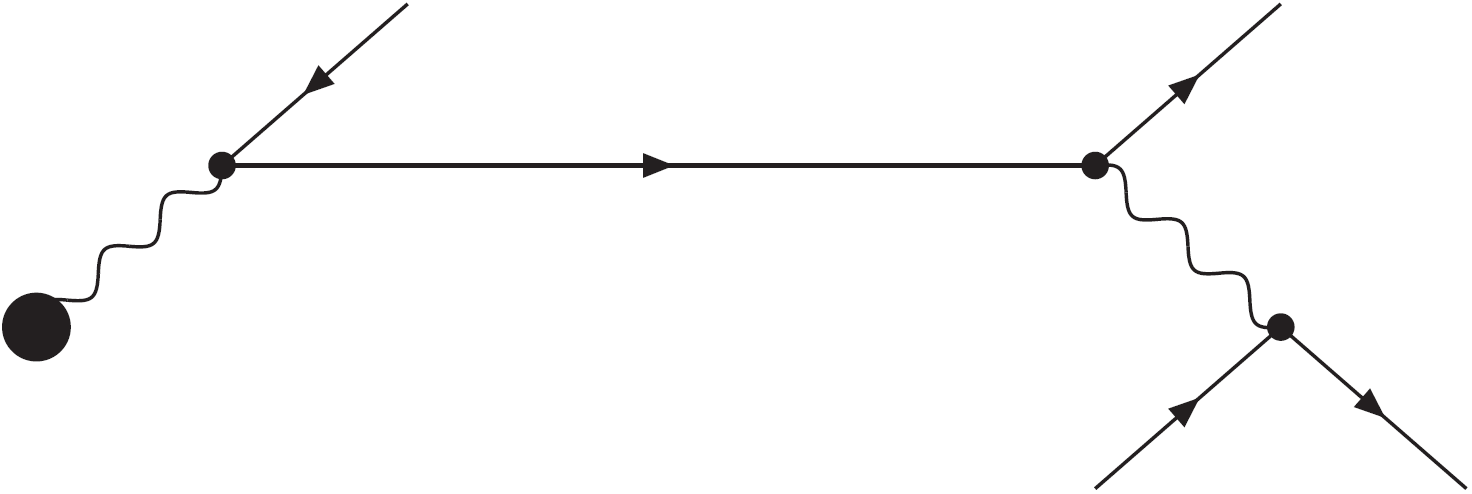}
\end{center}
\end{figure}
\vspace*{-3.52cm}
\begin{center}
\begin{picture}(193,81)(35,0)
\Text(70.0,94.0)[l]{$e^+ (q)$}\ArrowLine(67.5,88.0)(40.5,64.5)
\Text(33.5,65.5)[r]{$x$} \Photon(13.5,41.0)(40.5,64.5){2}{3.0}
\Text(53.5,48.5)[r]{$W^+$} \Vertex (13.5,41.0){5} \Vertex
(40.5,64.5){2} \ArrowLine(40.5,64.5)(167.5,64.5) \Vertex
(167.5,64.5){2} \Text(104.8,70.5)[b]{$\nu_k ( p_{\rm n} )$}
\ArrowLine(167.5,64.5)(194.5,88.0) \Text(197.5,94.0)[l]{$e^- ( k )$}
\Text(175.0,64.5)[l]{$y$} \Text(160.0,48.5)[l]{$W^+$}
\Photon(167.5,64.5)(194.5,41.0){2}{3.0} \Vertex (194.5,41.0){2}
\ArrowLine(167.5,17.5)(194.5,41.0) \Text(132.5,17.5)[l]{$e^- ( k_1 )$}
\ArrowLine(194.5,41.0)(221.5,17.5) \Text(225.0,17.5)[l]{$\nu_i ( k_2 )$}
\Text(363.0,60.5)[b]{\addtocounter{equation}{1}(\arabic{equation})}
\setcounter{diag2_CC}{\value{equation}}
\end{picture}
\end{center}
\vspace{0.1cm}
The amplitude corresponding to diagram (\arabic{diag2_CC}) should
be summed over the type $k$ of the intermediate neutrino mass
eigenstate, because all of them contribute. Since only the final
electron is detected in the experiment, the probability of the
process with $i$th neutrino mass eigenstate in the final state
should be summed over $i$ to give the probability of registering
an electron.

In order to use the foregoing formulas without
redefinitions, we retain the previous notations for  particles' 4-momenta
and nuclear values. Besides, we introduce the missing notations
$k_1$ and $k_2$ for the 4-momenta of the incoming electron and
outgoing neutrino $\nu_i$, respectively.

Using the approximation of Fermi's interaction and
distance-dependent propagator (\ref{dist_dep_prop_field}), keeping
the neutrino masses only in the exponential, one can write down
the amplitude corresponding to neutral-current diagram
(\arabic{diag2_NC}) in the momentum representation as follows:
\begin{equation}
\begin{split}
& M_i^{{\rm nc}}  = {\rm i}\frac{{G_{\rm F}^2 }}{{8\left| {\vec p_{\rm n} } \right|}} U_{1i}^ * \,  j_\mu ^{(1)} \left( {\vec P^{(1)} ,\vec P^{(1')} } \right) \bar \nu _i \left( {\vec k_2 } \right)\left( {1 + \gamma ^5 } \right)\gamma ^\rho \, \hat p_{\rm n} \times  \\
& \hphantom{M_i^{{\rm nc}}  =}
\times \left[ {\left( {1 - {\rm i}\vec j\vec \gamma } \right){\rm e}^{{\rm i}\frac{{p_{\rm n}^2  - m_i^2  + 2\mu _0 m_i \left| {\vec p_{\rm n} } \right|\overline H }}{{2\left| {\vec p_{\rm n} } \right|}}L}  + \left( {1 + {\rm i}\vec j\vec \gamma } \right){\rm e}^{{\rm i}\frac{{p_{\rm n}^2  - m_i^2  - 2\mu _0 m_i \left| {\vec p_{\rm n} } \right|\overline H }}{{2\left| {\vec p_{\rm n} } \right|}}L} } \right] \gamma ^\mu  \left( {1 - \gamma ^5 } \right)v\left( {\vec q} \right) \times  \\
& \hphantom{M_i^{{\rm nc}}  =} \times \left[ { \left( { -
\frac{1}{2} + \sin ^2 \theta _{\rm w} } \right)\bar u\left( {\vec
k} \right)\gamma _\rho  \left( {1 - \gamma ^5 } \right)u\left(
{\vec k_1 } \right) + \sin ^2 \theta _{\rm w} \, \bar u\left(
{\vec k} \right)\gamma _\rho  \left( {1 + \gamma ^5 }
\right)u\left( {\vec k_1 } \right) } \right] .
\end{split}
\end{equation}
The amplitude corresponding to charged-current diagram
(\arabic{diag2_CC}) summed over the index $k$ reads
\begin{equation}
\begin{split}
& M_i^{{\rm cc}}  =  - {\rm i}\frac{{G_{\rm F}^2 }}{{8\left| {\vec p_{\rm n} } \right|}}U_{1i}^ * \, j_\mu ^{(1)} \left( {\vec P^{(1)} ,\vec P^{(1')} } \right)\bar u\left( {\vec k} \right)\left( {1 + \gamma ^5 } \right)\gamma ^\rho \, \hat p_{\rm n}  \times  \\
& \hphantom{M_i^{{\rm cc}}  =} \times \sum\limits_{k = 1}^3
{\left| {U_{1k} } \right|^2 \left[ {\left( {1 - {\rm i}\vec j\vec
\gamma } \right){\rm e}^{{\rm i}\frac{{p_{\rm n}^{\rm 2}  - m_k^2
+ 2\mu _0 m_k \left| {\vec p_{\rm n} } \right|\overline H
}}{{2\left| {\vec p_{\rm n} } \right|}}L}  + \left( {1 + {\rm
i}\vec j\vec \gamma } \right){\rm e}^{{\rm i}\frac{{p_{\rm n}^{\rm
2}  - m_k^2 - 2\mu _0
m_k \left| {\vec p_{\rm n} } \right|\overline H }}{{2\left| {\vec p_{\rm n} } \right|}}L} } \right]}  \times  \\
& \hphantom{M_i^{{\rm cc}}  =} \times \gamma ^\mu  \left( {1 -
\gamma ^5 } \right)\upsilon \left( {\vec q} \right) \cdot \bar \nu
_i \left( {\vec k_2 } \right)\gamma _\rho  \left( {1 - \gamma ^5 }
\right)u\left( {\vec k_1 } \right).
\end{split}
\end{equation}

The squared modulus of the total amplitude $M_i  = M^{\rm nc}_i  +
M^{\rm cc}_i$, averaged  and summed over particles' polarizations,
factorizes in the approximation $p_{\rm n}^2=0$. Following the
procedure described above and summing the differential probability
over the final neutrino type $i$, we arrive at the probability of
detecting an electron in the process at hand:
\begin{equation} \label{final_prob_NC}
\frac{{{\rm d}W}}{{{\rm d}\Omega }} = \int\limits_{\left| {\vec p}
\right|_{\min } }^{\left| {\vec p} \right|_{\max } } {\frac{{{\rm
d}^{\rm 3} W_{\rm P} }}{{{\rm d}^3 p}} \, W_{\rm D} \left(
{L,\overline H } \right)\left| {\vec p} \right|^2 {\rm d} | {\vec
p} |} .
\end{equation}
Here the differential production probability ${{{{\rm d}^3 W_{\rm
P}} \mathord{\left/ {\vphantom {{{\rm d}^3 W_{\rm P}} {{\rm d}^3
p}}} \right. \kern-\nulldelimiterspace} {{\rm d}^3 p}}}$ is given
by (\ref{decay_prob}) and the total detection probability $W_{\rm
D} \left( {L,\overline H } \right)$, which is the sum of the
detection probabilities for each neutrino mass eigenstate in the
final state, reads
\begin{equation} \label{det_prob_NC}
\begin{split}
& W_{\rm D} \left( {L,\overline H } \right) = \frac{{G_{\rm F}^2 m_e }}{{2\pi }}\left\{ {\left[ {\left( {1 - 4\sin ^2 \theta _{\rm w}  + 8\sin ^4 \theta _{\rm w} } \right)\Delta T - 4\sin ^2 \theta _{\rm w} \left( {\frac{{\sin ^2 \theta _{\rm w} }}{{T_{\max } }} - \frac{{m_e }}{{4\left| {\vec p} \right|^2 }}} \right)\Delta T^2 \, + } \right.} \right. \\
& \hphantom{W_{\rm D} \left( {L,\overline H } \right) =
\frac{{G_{\rm F}^2 m_e }}{{2\pi }} \bigg\{ \bigg[}
\left. { + \, \frac{4}{3} \frac{\sin ^4 \theta _{\rm w}}{{\left| {\vec p} \right|^2 }}\Delta T^3 } \right]\sum\limits_{i = 1}^3 {\left| {U_{1i} } \right|^2 \cos ^2 \left( {\mu _0 m_i \overline H L} \right) + }  \\
& \hphantom{W_{\rm D} \left( {L,\overline H } \right) =
\frac{{G_{\rm F}^2 m_e }}{{2\pi }} \bigg\{ } \left. { + \, 8\sin ^2
\theta _{\rm w} \left[ {\Delta T - \frac{{m_e }}{{4\left| {\vec p}
\right|^2 }}\Delta T^2 } \right]P_{ee} \left( {\left| {\vec p}
\right|,L,\overline H } \right)} \right\} ,
\end{split}
\end{equation}
where
\begin{equation}
\Delta T \equiv T_{\max }  - T_{\min } \, , \qquad \Delta T^2
\equiv T_{\max }^2  - T_{\min }^2 \, , \qquad \Delta T^3  \equiv
T_{\max }^3  - T_{\min }^3 \, ,
\end{equation}
\begin{equation} \label{T_max}
T_{\max }  = \frac{{2\vec p^{\,2} }}{{2\left| {\vec p} \right| +
m_e }}.
\end{equation}
Here $T_{\max }$  is the maximum kinetic energy of the final
electron in the detection process, determined by energy-momentum
conservation, and $T_{\min }$ is the minimum kinetic energy of the
electron accessible for registrations by a specific detector. The
lower integration limit $\left| {\vec p} \right|_{\min }$ in
(\ref{final_prob_NC}) is connected with $T_{\min }$ by a relation
similar to (\ref{T_max}), resolving which we get
\begin{equation}
\left| {\vec p} \right|_{\min }  = \frac{1}{2}\left( {T_{\min }  +
\sqrt {T_{\min } \left( {T_{\min }  + 2m_e } \right)} } \right) .
\end{equation}

It is easy to verify that the detection probability
(\ref{det_prob_NC}) is expressed through the oscillating factor
(\ref{prob_osc}) and the Standard Model neutrino scattering
probabilities $W _{\nu _e e}$ and $W _{\nu _\mu e}$, calculated
for the same values of $T_{\min }$ and $T_{\max }$, by the
relation
\begin{equation}\label{prob_NC}
W_{\rm D}  = P_{ee} \left( {\left| {\vec p} \right|,L,\overline H
} \right) {W _{\nu _e e}\left({\left| {\vec p} \right|}\right)}   + \left(
\sum\limits_{i = 1}^3 {\left| {U_{1i} } \right|^2 \cos ^2 \left(
{\mu _0 m_i \overline H L} \right)}  - P_{ee} \left( {\left| {\vec
p} \right|,L,\overline H } \right) \right) W _{\nu _\mu
e}\left({\left| {\vec p} \right|}\right),
\end{equation}
which  differs from the standard one in vacuum by an extra
field-dependent  oscillating factor taking into account the
neutrino spin rotation.

\section{Specific examples}

Let us discuss some examples, where neutrinos are produced  in the
decay of ${^{15} {\rm O}}$
\begin{equation} \label{O15}
    {^{15} {\rm O}} \to {^{15} {\rm N}} + e^ +   + \nu_i
\end{equation}
or in the electron capture reaction
\begin{equation} \label{Be7}
    {^{7} {\rm Be}} + e^ - \to {^{7} {\rm Li}}  + \nu_i
\end{equation}
and detected by a gallium-germanium detector in the reaction
\begin{equation} \label{Detectors}
    \nu_i +
   {^{71} {\rm Ga}}  \\
 \to    {^{71} {\rm Ge}}   + e^- .
\end{equation}
Both reactions contribute to the solar neutrino flux, the first
reaction having a wide energy spectrum and the second one
producing almost monoenergetic neutrinos.

We start with the ${^{15} {\rm O}}$-source.
Nuclear reactions (\ref{O15}), (\ref{Detectors}) are allowed
transitions \cite{Bohr-Mottelson},  which means that we can
neglect the nucleon positions and momenta and consider the
interaction or decay of a nucleon as if it were at rest. Thus,
neglecting the dependence of the nuclear form-factors on the
momentum transfer \cite{Bohr-Mottelson} and also neglecting the
possible contribution of the excited states of the final nuclei,
one can approximate the product of the differential probability of
neutrino production and the probability of neutrino detection by
the function
\begin{eqnarray}
\begin{split}
\frac{{{\rm d}^3 W_{\rm P} }}{{{\rm d}^3 p}} \, W_{\rm D}  = & \ C\sqrt {\left( {\left| {\vec p} \right|_{\max }  - \left| {\vec p} \right|} \right)\left( {\left| {\vec p} \right|_{\max }  - \left| {\vec p} \right| + 2m_e} \right)} \left( {\left| {\vec p} \right|_{\max }  - \left| {\vec p} \right| + m_e} \right) \times \\
&\times \sqrt {\left( {\left| {\vec p} \right| - \left| {\vec p} \right|_{\min } } \right)\left( {\left| {\vec p} \right| - \left| {\vec p} \right|_{\min }  + 2m_e} \right)} \left( {\left| {\vec p} \right| - \left| {\vec p} \right|_{\min }  + m_e} \right).
\end{split}
\end{eqnarray}
Here $C$ is a constant, the explicit form  of which is unimportant
for us, because we will normalize the probability
(\ref{final_prob}) so that it equals unity at the point $L=0$;
$\left| {\vec p} \right|_{\max }$ and $\left| {\vec p}
\right|_{\min }$ are the same as the integration limits in
(\ref{final_prob}), determined by (\ref{p_min_max}). For the
production and detection processes under consideration we have:
\begin{equation} \label{ps}
\left| {\vec p} \right|_{\max } = 1732 {\ \rm keV,}\quad   \left|
{\vec p} \right|_{\min } = 232 {\ \rm keV}.
\end{equation}

\begin{figure}[h!!!]
\begin{minipage}[h]{0.49\linewidth}
\begin{center}
\includegraphics[width=1\linewidth]{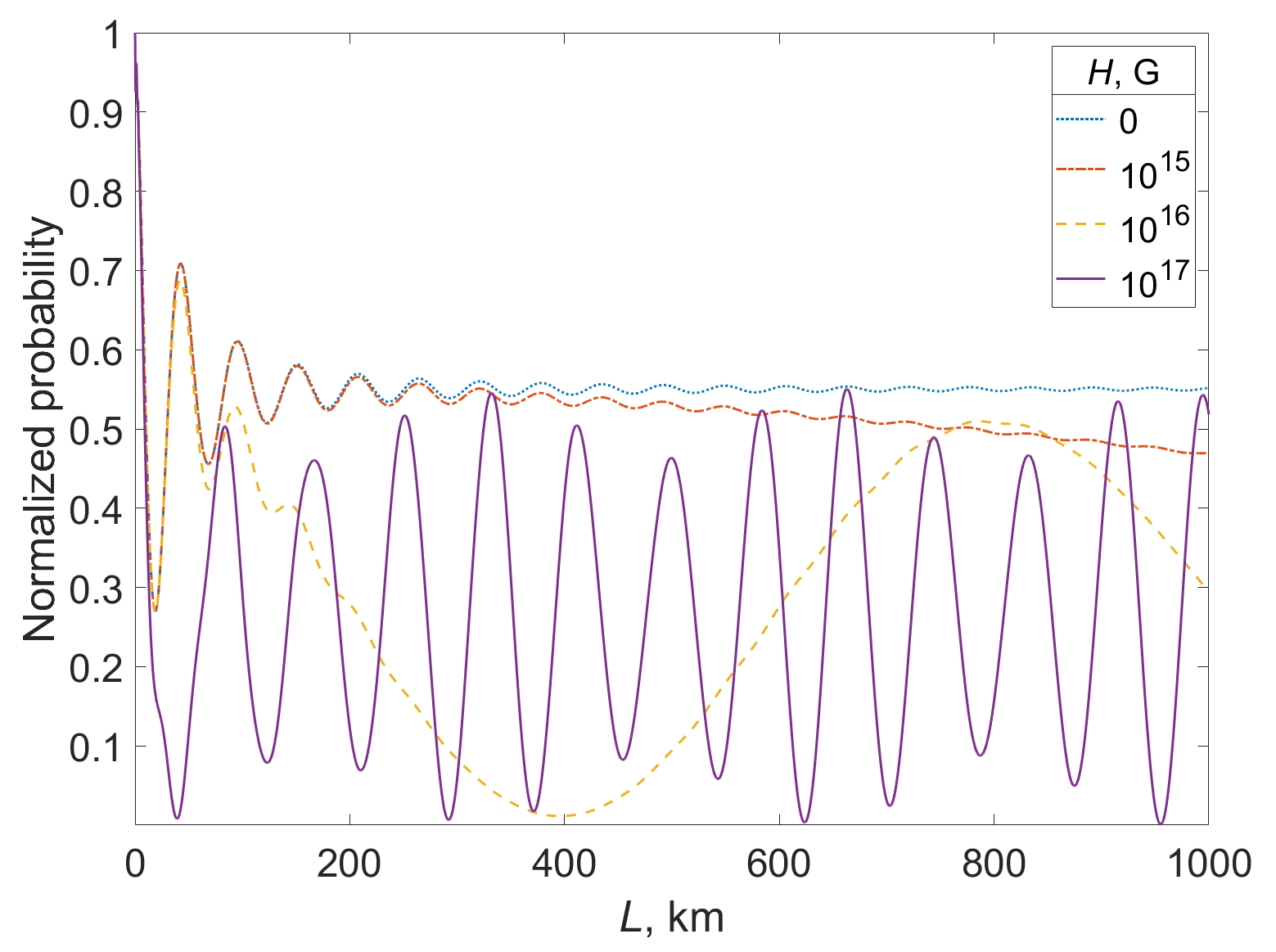}
\\ a) The first set of masses (\ref{set1}).
\end{center}
\end{minipage}
\hfill
\begin{minipage}[h]{0.49\linewidth}
\begin{center}
\includegraphics[width=1\linewidth]{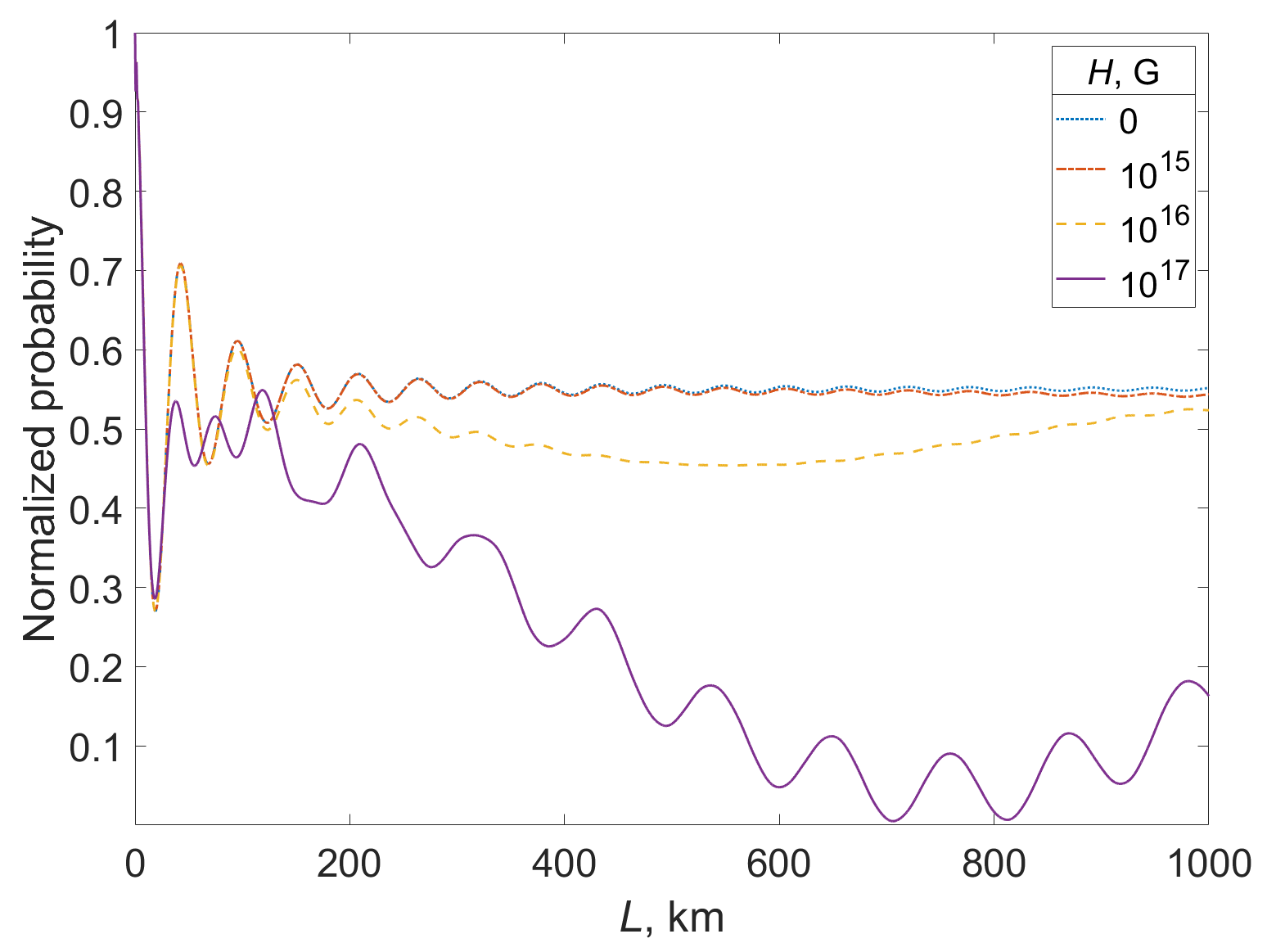}
\\ b) The second set of masses (\ref{set2}).
\end{center}
\end{minipage}
\caption{Normalized probabilities (\ref{final_prob}) of the
neutrino oscillation  processes in constant homogeneous magnetic
field of various magnitude; neutrino production in the ${^{15}
{\rm O}}$ decay and detection by a Ga-Ge detector.}
\label{osc_O15_GaGe}
\end{figure}

Below the following values of the mixing angles are used \cite{Zyla:2020zbs}:
\begin{equation}
 \sin ^2  {\theta _{12} } = 0.307, \qquad \sin ^2 {\theta _{23} } = 0.545, \qquad \sin ^2 {\theta _{13} } = 2.18 \cdot 10^{ - 2}.
\end{equation}
Let the constant $\mu_0$  take the value predicted by the Standard
Model, $\mu _0  = {{3eG_{\rm F} } \mathord{\left/ {\vphantom
{{3eG_{\rm F} } {8\sqrt 2 \pi ^2 }}} \right.
\kern-\nulldelimiterspace} {8\sqrt 2 \pi ^2 }} = 9.488 \cdot
10^{-26} {\ \rm eV^{-2}}$. Taking into account the experimental
restrictions for the masses of the normally ordered neutrinos
\cite{Zyla:2020zbs}, which is the most likely scenario,
\begin{equation}
 \Delta m_{21}^2  = 7.53 \cdot 10^{ - 5} {\ \rm  eV}^2 , \qquad \Delta m_{32}^2  = 2.45 \cdot 10^{ - 3} {\ \rm  eV}^2,
\end{equation}
and the cosmological limit \cite{Aghanim:2018eyx}
\begin{equation}
    m_1  + m_2  + m_3  < 0.120 {\ \rm eV},
\end{equation}
we first take one set of masses,
\begin{equation} \label{set1}
m_1 = 0.0114{\ \rm eV}, \qquad m_2 = 0.0143{\ \rm eV}, \qquad m_3 = 0.0515{\ \rm eV},
\end{equation}
and then the second one,
\begin{equation} \label{set2}
m_1 = 6 \cdot 10^{ - 4} {\ \rm eV}, \qquad m_2 = 8.7 \cdot 10^{ - 3}{\ \rm eV}, \qquad m_3 = 0.0503{\ \rm eV}.
\end{equation}
The results of numerical integration of formula (\ref{final_prob})
in the case of homogeneous magnetic field with these parameters
for the two sets of masses, in the absence of external magnetic
field and for its transverse values $10^{15}$, $10^{16}$,
$10^{17}$ G are presented in Fig.~\ref{osc_O15_GaGe}.
Such huge magnitudes of the magnetic field, close to  the
magnetism limit of neutron stars $10^{18}$ G, are taken just for
an illustration, in order to show both the momentum-dependent and
field-dependent oscillations in one picture. As it was discussed
in \cite{Egorov:2019vqv}, the momentum-dependent oscillations fade
out at a rather short distance from the source due to the presence
of the neutrino momentum spread, and at larger distances only the
field-dependent oscillations are basically left.

\begin{figure}[h]
\begin{minipage}[h]{0.49\linewidth}
\begin{center}
\includegraphics[width=1\linewidth]{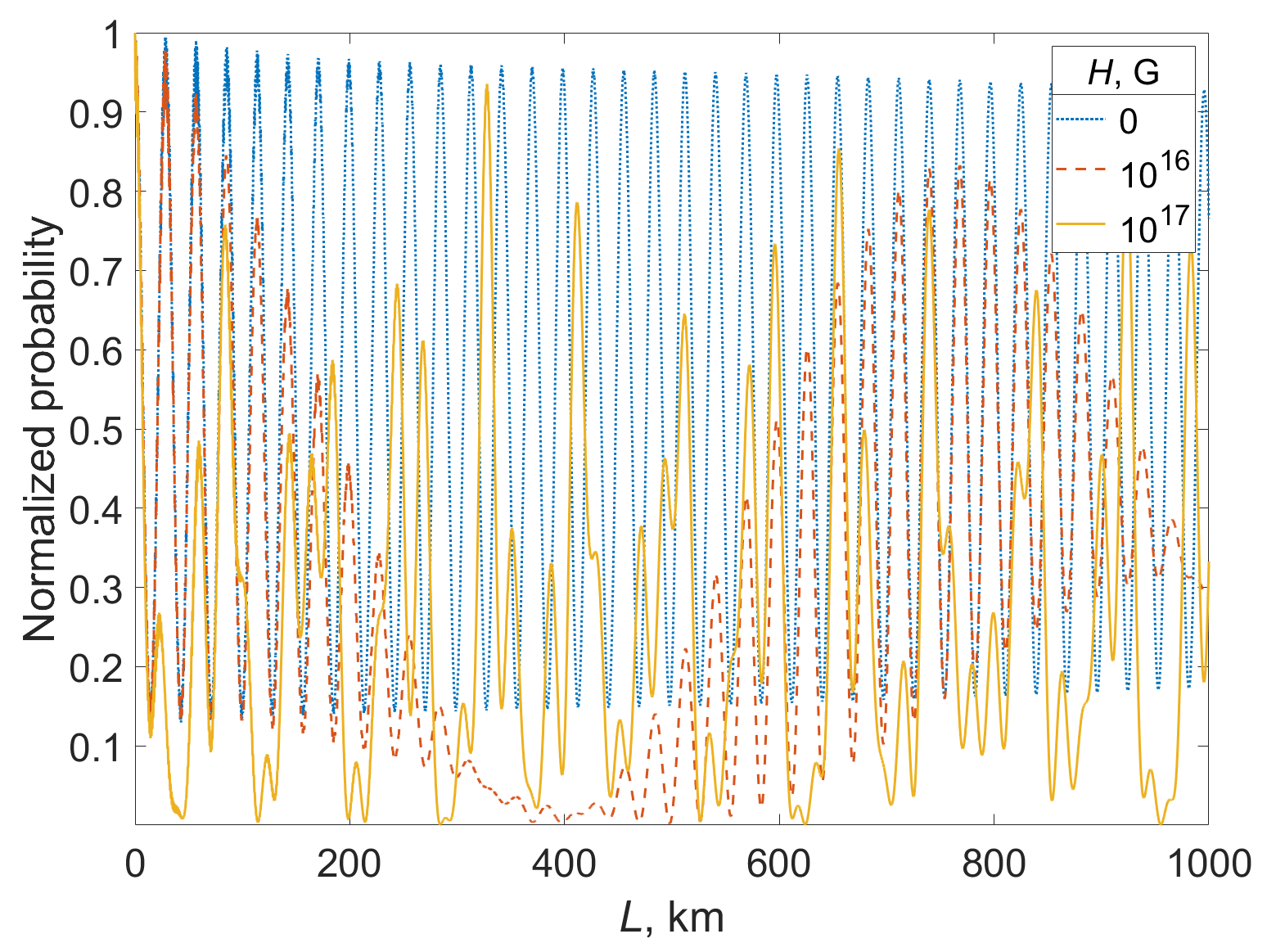}
\\ a) The first set of masses (\ref{set1}).
\end{center}
\end{minipage}
\hfill
\begin{minipage}[h]{0.49\linewidth}
\begin{center}
\includegraphics[width=1\linewidth]{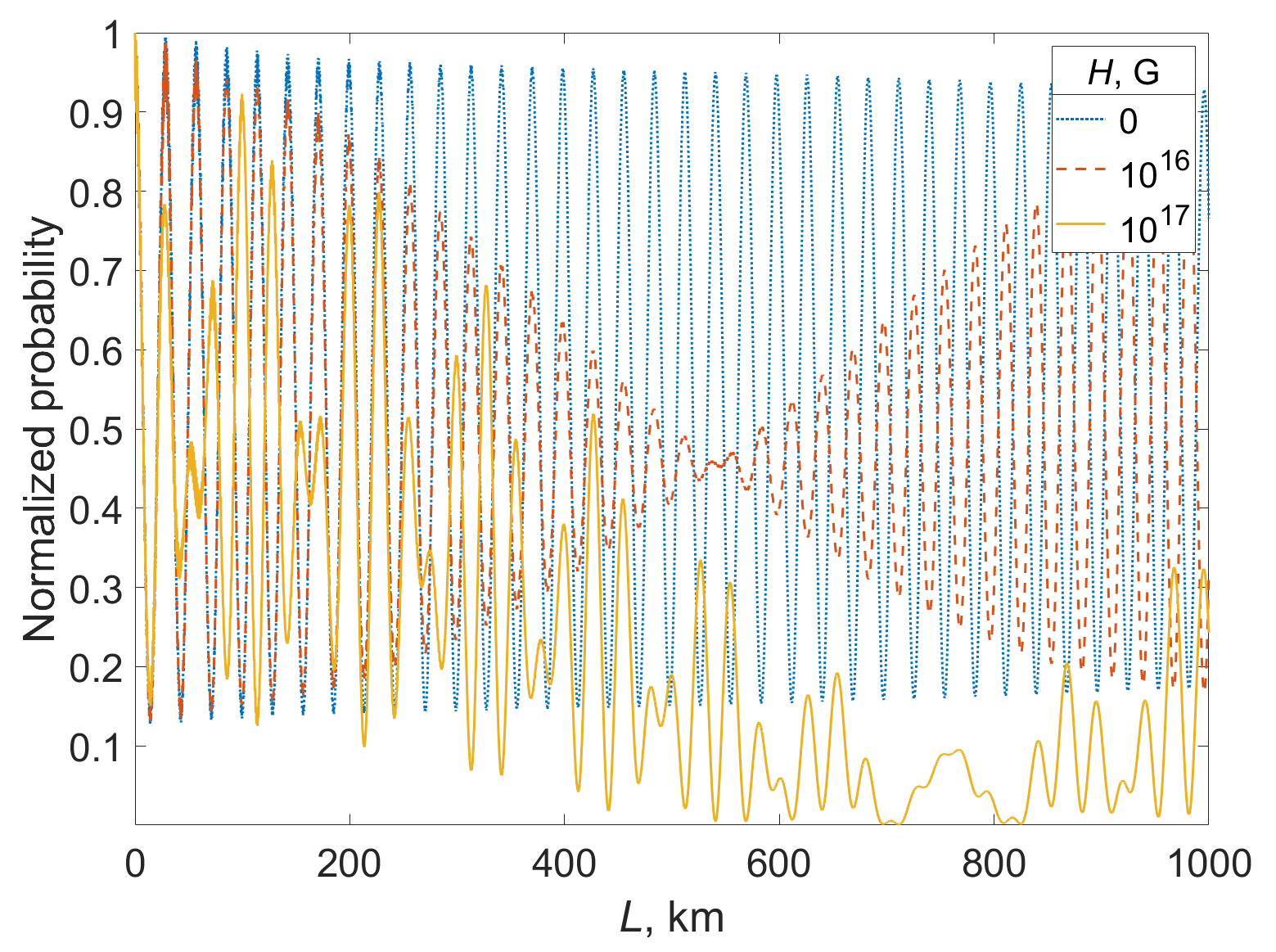}
\\ b) The second set of masses (\ref{set2}).
\end{center}
\end{minipage}
\caption{Averaged normalized probabilities (\ref{aver_prob_Be}) of
the neutrino oscillation processes in constant homogeneous
magnetic field of various magnitude; neutrino production in the
${^{7} {\rm Be}}$-electron interaction and detection by a Ga-Ge
detector.} \label{osc_Be7_GaGe}
\end{figure}

Now let us consider a source, where  reaction (\ref{Be7}), $ {^{7}
{\rm Be}} + e^ - \to {^{7} {\rm Li}}  + \nu_i$, takes place. This
reaction also belongs to the allowed transitions, and one can
neglect nuclear form-factors in calculating the matrix elements.
In the non-relativistic approximation the squared modulus of the
production amplitude, averaged and summed over the polarizations
of the nuclei and particles, can be written as \cite{Okun1985}
\begin{equation}
\left\langle {\left| {M_{\rm P} } \right|^2 } \right\rangle  = q^0
p_{\rm n}^0 \left( {C_0  + C_1 \vec v_e \vec v_\nu } \right),
\end{equation}
where $C_{0,1}$ are constants, $q^0$ is the initial electron
energy, $p_{\rm n}^0$ is the final neutrino energy and $\vec v_e$,
$\vec v_\nu$ are the electron and neutrino velocities,
respectively. Reaction (\ref{Be7}), having a two-particle final
state, produces neutrinos with  a fixed energy, when the initial
particles have definite energies. The neutrino energy has only a
thermal broadening due to the spread in the energies of the
initial particles. Since the electron mass is 4 orders of
magnitude smaller then the ${^{7} {\rm Be}}$ nucleus mass, it is
the electron energy spread that mainly contributes to the neutrino
energy broadening. Taking the plasma temperature to be $T = 1.5
\cdot 10^7$ K, as in the center of the Sun, which is consistent
with the non-relativistic approximation, we average probability
(\ref{final_prob_2body}) with the Maxwell-Boltzmann distribution
for the initial electron momentum. As a result, the term with the
velocities vanishes, and, neglecting the exited states of ${^{7}
{\rm Li}}$ nucleus, we arrive at the expression
\begin{equation} \label{aver_prob_Be}
\begin{split}
 & \overline {\frac{{{\rm d}W}}{{{\rm d}\Omega }}}  = C\int\limits_\Delta ^\infty  {\sqrt {\left( {\left| {\vec p} \right| - \left| {\vec p} \right|_{\min } } \right)\left( {\left| {\vec p} \right| - \left| {\vec p} \right|_{\min }  + 2m_e } \right)} \left( {\left| {\vec p} \right| - \left| {\vec p} \right|_{\min }  + m_e } \right) \times }  \\
 & \hspace{6cm}
\times {\rm e}^ { - \frac{ \left| {\vec p} \right| -\Delta
}{{kT}}} \sqrt {\left| {\vec p} \right| - \Delta } \, P_{ee}
\left( {\left| {\vec p} \right|,L,\overline H  } \right) \left|
{\vec p} \right|^2 {\rm d} |{\vec p}| ,
\end{split}
\end{equation}
where $\Delta  = M_{{\rm Be}}  + m_e  - M_{{\rm Li}} = 862$ keV
is the reaction energy release (the neutrino mass is neglected),
$\left| {\vec p} \right|_{\min }$ for a Ga-Ge detector is
presented in (\ref{ps}), $\left| {\vec p} \right|$ is the neutrino
momentum magnitude and $C$ is a constant. Performing the numerical
integration in the latter formula for two sets of masses
(\ref{set1})--(\ref{set2}) and a Ga-Ge detector, one arrives at
the results depicted in Fig.~\ref{osc_Be7_GaGe}.
We see, once again, that the narrower neutrino momentum
distribution leads to larger coherence lengths of the
momentum-dependent oscillations. Namely, in the case of ${^{7}
{\rm Be}} \, + \, e^-$ source of temperature $T = 1.5 \cdot 10^7$
K the coherence length turns out to be about 10\,000 km. We do not
present the corresponding figure, because the oscillations
actually merge into a band on this scale. The curves in
Fig.~\ref{osc_Be7_GaGe} b) resemble those obtained in
\cite{Chukhnova:2019oum} for the spin-flavor oscillations of
monoenergetic neutrinos in homogeneous magnetic field.

Borexino \cite{Borexino:2017fbd} and GEMMA \cite{Beda:2012zz}
experiments restrict the neutrino magnetic moment from above by
the value about $2.8 \cdot 10^{-11} \, \mu_{\rm B} = 8.3 \cdot
10^{- 18} {\ \rm eV^{-1}}$, where $\mu_{\rm B} = e\hbar
\mathord{\left/ {\vphantom {e\hbar {2m_e c }}} \right.
\kern-\nulldelimiterspace} {2m_e c }$ is the Bohr magneton. This
bound is 10 orders of magnitude larger than the value, predicted
with the Standard Model $\mu_0$, which allows 10 orders of
magnitude weaker magnetic field for the same effect.

The results depicted in Figs.~\ref{osc_O15_GaGe},
\ref{osc_Be7_GaGe} refer to the case of a homogeneous magnetic
field present along the entire neutrino path. If the neutrinos
travel in magnetic field only a part of the path, for fairly large
values of $L$, where the momentum-dependent oscillations fade out,
the normalized probabilities of all processes are defined only by
the magnetic field and  go to the asymptotic value
\begin{equation}\label{asymp}
W_{\rm asym} = \sum\limits_{i = 1}^3 \left| {U_{1i} } \right|^4 -
\sum\limits_{i = 1}^3 {\left| {U_{1i} } \right|^4 \sin^2
\delta_i }\,, \quad \delta_i  = \mu_0 m_i \int\limits_{D} H \left({ l }\right) {\rm d}l \,,
\end{equation}
where $ \sum\nolimits_{i = 1}^3 \left| {U_{1i} } \right|^4 \simeq
0.550 $ is the  asymptotic value in vacuum,  $\delta_i$ is the
phase accumulated by the neutrino mass eigenstate $\nu_i$ on its path
and $D$ is the field region with respect to coordinate $l$ along the neutrino trajectory.

\begin{figure}[htb]
\begin{minipage}[h]{0.520\linewidth}
\begin{center}
\includegraphics[width=1\linewidth]{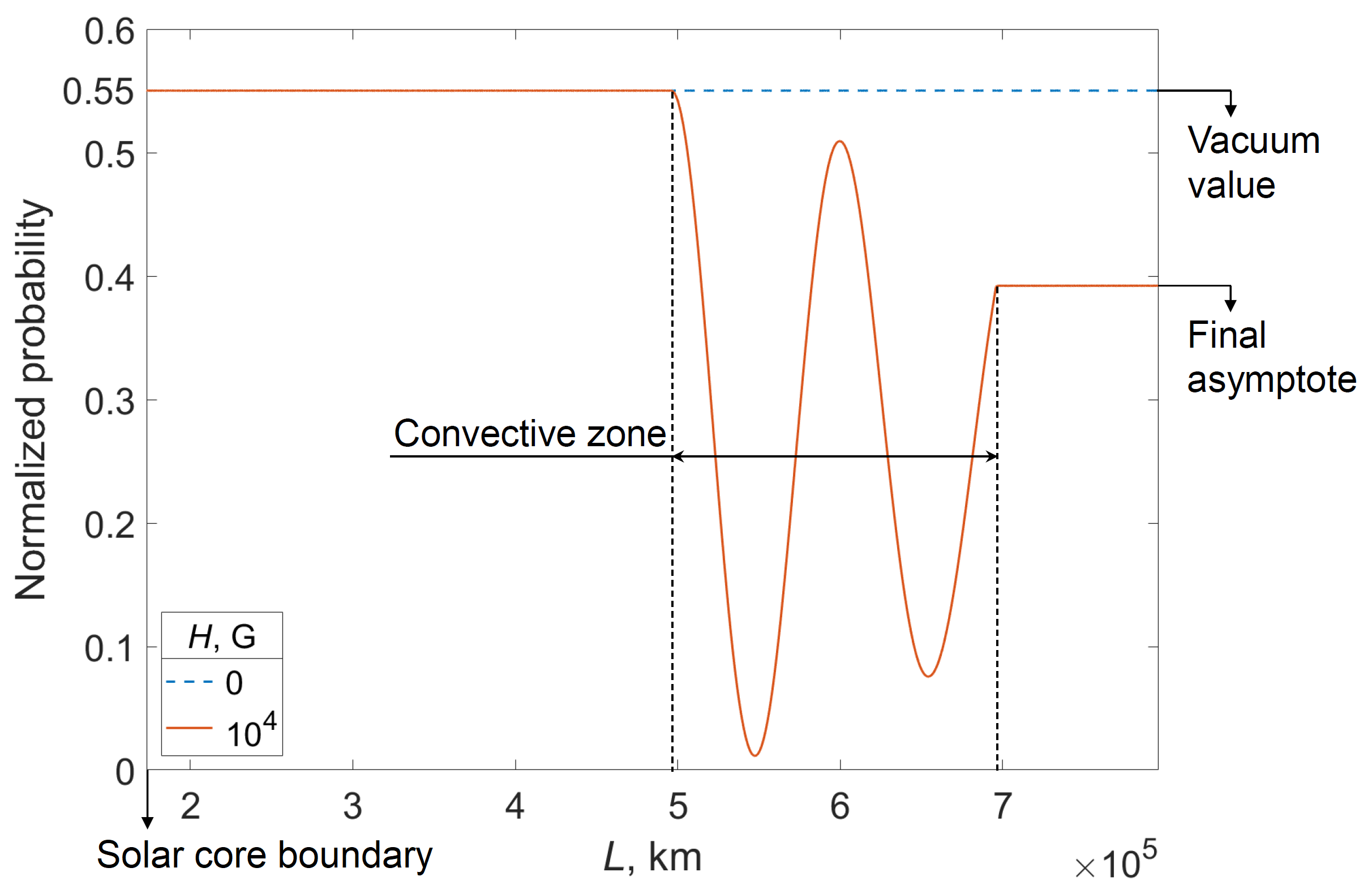}
\end{center}
\caption{Normalized probability of the neutrino oscillation
process in the convective zone of the Sun.} \label{osc_conv}
\end{minipage}
\hfill
\begin{minipage}[h]{0.445\linewidth}
\begin{center}
\includegraphics[width=1\linewidth]{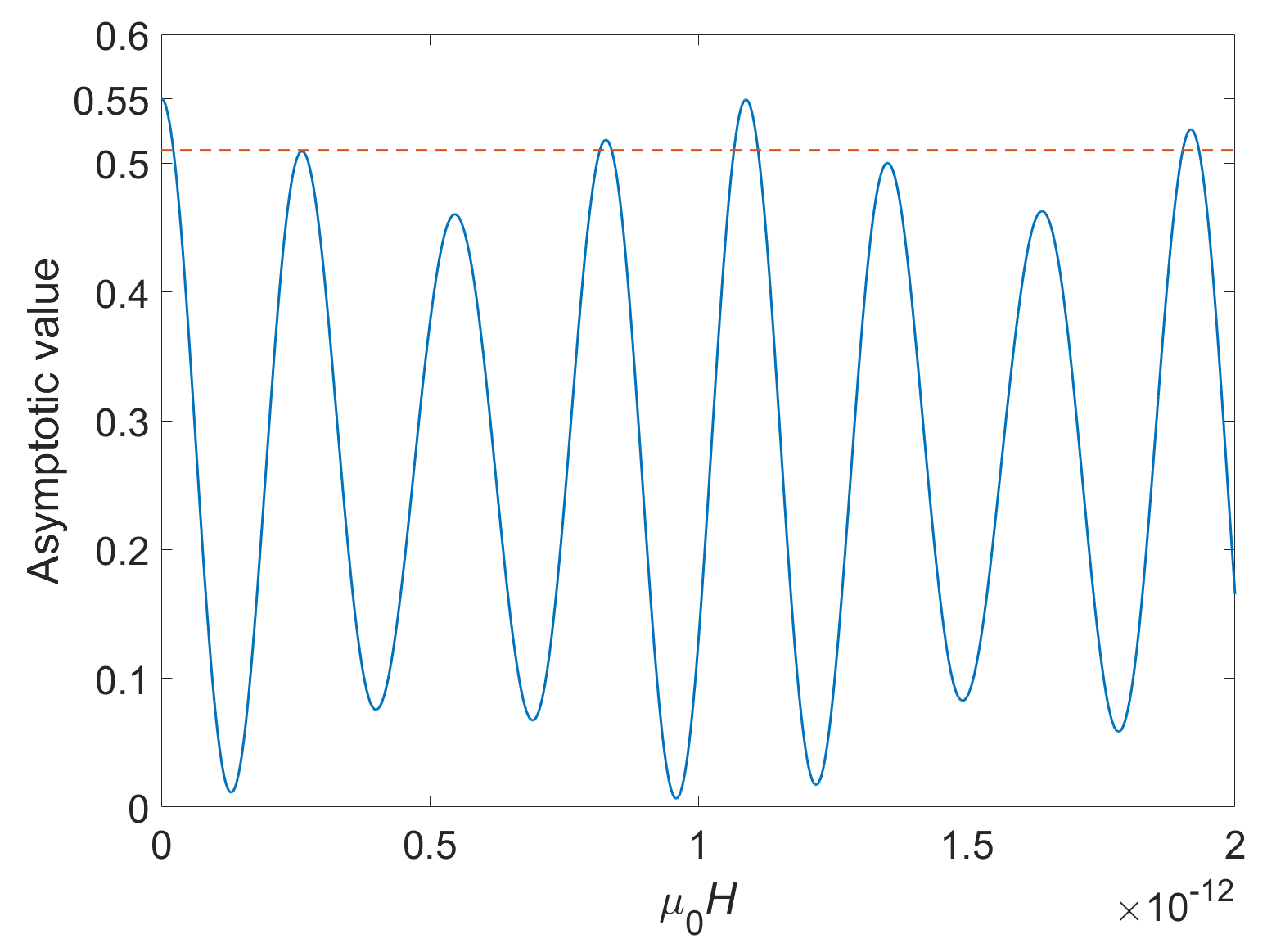}
\end{center}
\caption{Asymptotic value (\ref{as_Sol}) as a function of $\mu _0 \overline H$. \textcolor{white}{Text for easy picture alignment.}}
\label{plot_asymp}
\end{minipage}
\end{figure}

Let us consider, for example, the solar neutrinos, which are
produced in the solar core. As we have shown, even for
monoenergetic neutrino sources the coherence length is of the
order of 10\,000 km. Therefore, when the neutrinos come to the
convective zone, all the momentum-dependent oscillations fade out,
and only the oscillations due to the solar magnetic field present
in this zone remain. These oscillations for $\mu_1 = 2.8 \cdot
10^{-11} \, \mu_{\rm B}$, i.e., near the upper limit set by the
Borexino and GEMMA experiments
\cite{Borexino:2017fbd,Beda:2012zz}, the solar magnetic field
$10^4$ G used in paper \cite{Okun1986}, and the first mass set
(\ref{set1}) are shown in Fig.~\ref{osc_conv}, where the
asymptotic value is given by
\begin{equation}\label{as_Sol}
W_{\rm asym}^{\rm sol} = \sum\limits_{i = 1}^3 \left| {U_{1i} } \right|^4 -
\sum\limits_{i = 1}^3 {\left| {U_{1i} } \right|^4 \sin^2 \left(
\mu_0 m_i \overline H L_{\rm conv}\right)}, \quad \overline H L_{\rm conv}
=  \int\limits_{\scriptstyle {\rm convective} \atop
  \scriptstyle {\rm zone}} H \left({ l }\right) {\rm d}l \, ,
\end{equation}
with $\mu_0 \overline H \simeq 5 \cdot 10^{-13}$ and $L_{\rm conv}
\simeq 200\,000$ km. We note that, for the neutrino energy above
200 {\ \rm keV}, the size of the field region  $L_{\rm conv}$ and
the chosen value of the neutrino magnetic moment, adiabaticity
condition (\ref{adiabat}) is fulfilled with a great accuracy.
Since the thickness of the solar convective zone $L_{\rm conv}$ is
fixed and the product $\mu_0 \overline H$ varies depending on the
model, it is useful to plot the asymptotic value of the normalized
probability as a function of this product, which is shown in
Fig.~\ref{plot_asymp} (again, the first neutrino mass set is
chosen).

The normalized probability discussed in this paper can be
considered as the ratio of the number of  neutrinos detected in
the presence of oscillations to the number of neutrinos that would
have been detected in the absence of oscillations. For the
processes with only charged-current interaction its asymptotic
value does not depend on the production and detection processes,
and for this reason it gives a theoretical prediction for the
ratio of the flux of solar neutrinos  measured in an experiment
with a Ga-Ge (or Cl-Ar) detector to that predicted by the standard
solar model. This ratio for the GALLEX + GNO experiments is $0.58
\pm 0.07$ and for the SAGE experiment $0.59 \pm 0.07$
\cite{Bahcall:2000nu}, which is well consistent with the curve in
Fig.~4. We see that there are many values of
$\mu_0 \overline H$ giving the asymptotic value above the lower
experimental limit (dotted line). This means that, if we obtain an experimental
value of the neutrino magnetic moments, we will be able to obtain
restrictions on the solar magnetic field from the solar neutrino
experiments, and vice versa.

In the case of neutrino detection through both charged- and
neutral-current interaction the situation is different, because,
as it is clearly seen in formula (\ref{prob_NC}), the oscillating
factor $P_{ee} \left( {\left| {\vec p} \right|,L,\overline H }
\right)$  does not factorize. For this reason the asymptotic value
of the normalized probability depends not only on the production
process, but also on the energy range, in which the neutrinos are
detected. Here we consider the same production processes, the
$^{15}\rm O$ decay and the electron capture by ${^{7} {\rm Be}}$,
and the detection by a Cherenkov detector, which is capable of
measuring the neutrino energy.

\begin{figure}[h]
\begin{minipage}[h]{0.49\linewidth}
\begin{center}
\includegraphics[width=1\linewidth]{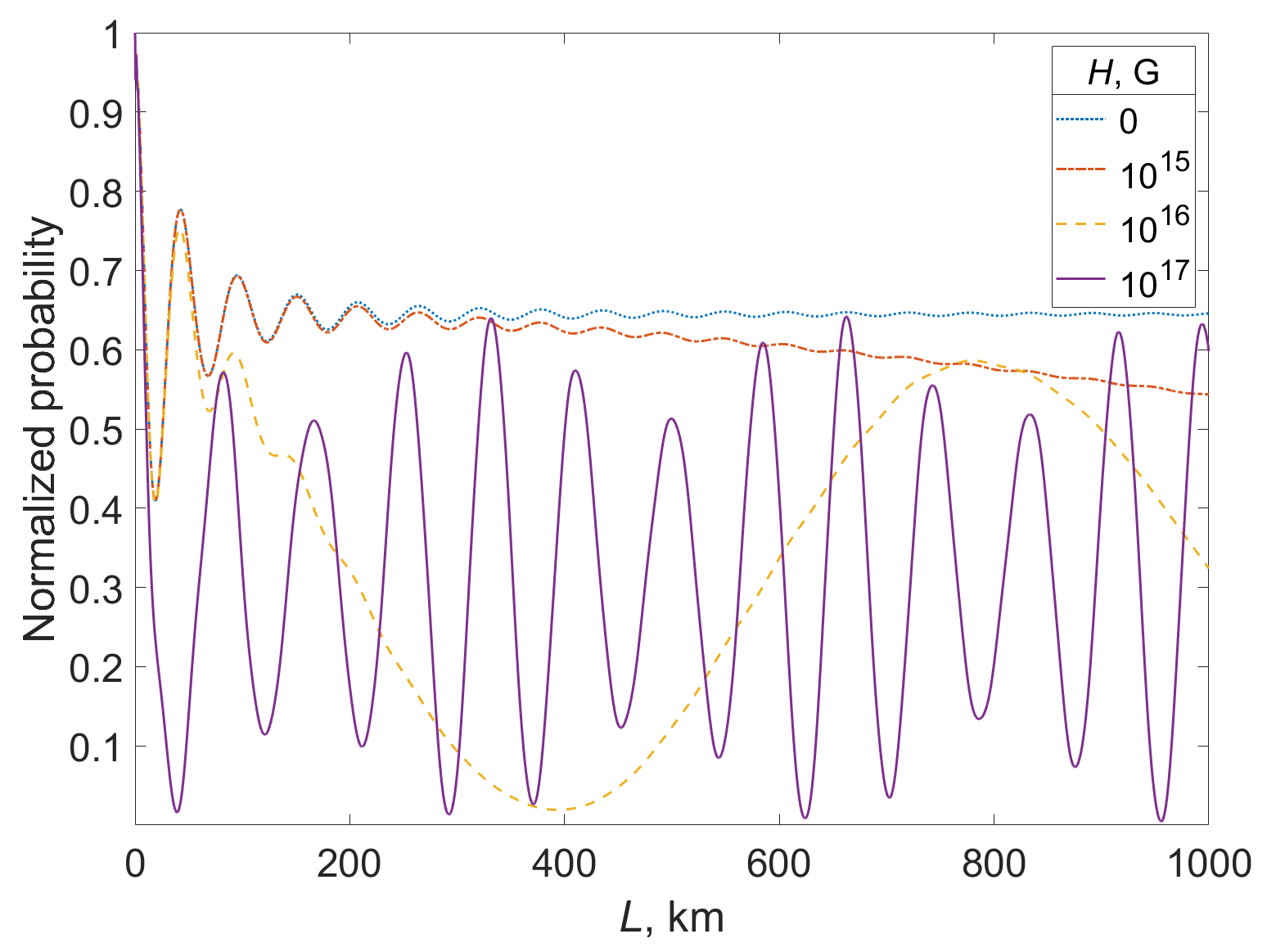}
\\ a) The first set of masses (\ref{set1}).
\end{center}
\end{minipage}
\hfill
\begin{minipage}[h]{0.49\linewidth}
\begin{center}
\includegraphics[width=1\linewidth]{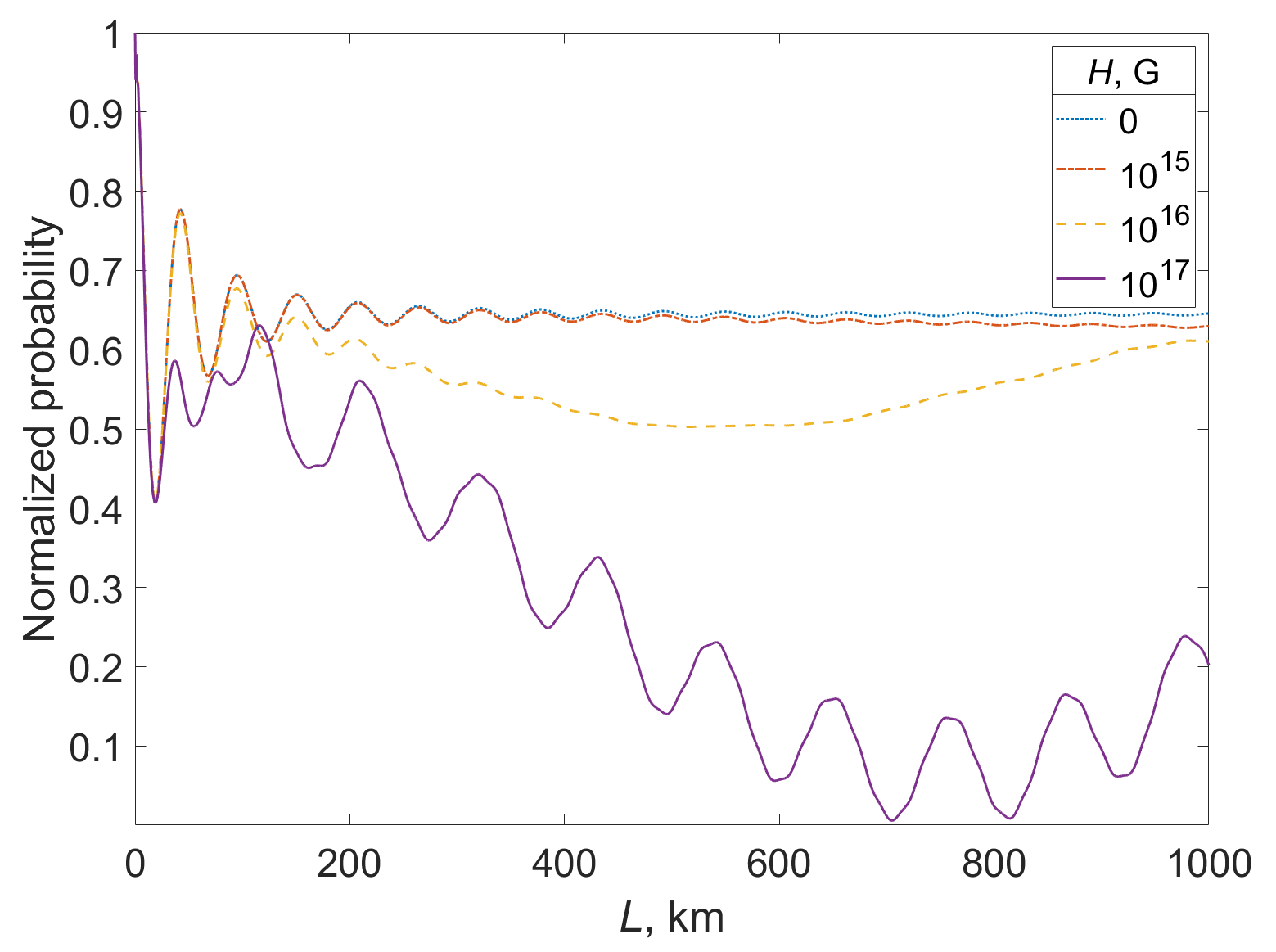}
\\ b) The second set of masses (\ref{set2}).
\end{center}
\end{minipage}
\caption{Normalized probabilities (\ref{final_prob_NC}) of the
neutrino oscillation processes  in constant homogeneous magnetic
field of various magnitude; neutrino production in the $^{15}\rm
O$ decay and detection by a water-based Cherenkov detector.}
\label{osc_O15_Cher}
\end{figure}

For  $^{15}\rm O$ decay (\ref{O15}), neglecting the dependence of
the nuclear form-factors on the momentum  transfer, the
differential probability of neutrino production can again be
approximated by the function
\begin{equation}
\frac{{{\rm d}^3 W_{\rm P} }}{{{\rm d}^3 p}} = C\sqrt {\left( {\left| {\vec p} \right|_{\max }  - \left| {\vec p} \right|} \right)\left( {\left| {\vec p} \right|_{\max }  - \left| {\vec p} \right| + 2m_e} \right)} \left( {\left| {\vec p} \right|_{\max }  - \left| {\vec p} \right| + m_e} \right),
\end{equation}
where the maximum neutrino momentum $\left| {\vec p} \right|_{\max
} = 1732$ keV. We take $\left| {\vec p} \right|_{\min } = 421$
keV, which corresponds to a water-based Cherenkov detector. The
results of numerical integration for the two neutrino mass sets
are depicted in Fig.~\ref{osc_O15_Cher}.

The asymptotic values for the oscillation processes with neutral
current can be found as follows. Substituting the asymptotic value
of $ P_{ee} \left( {\left| {\vec p_{\rm n} } \right|,L, \overline
H} \right)$ (\ref{prob_osc}) into formula (\ref{final_prob_NC})
for the probability and using the explicit form of $W_{\rm D}
\left( {L,\overline H } \right)$ (\ref{det_prob_NC}), we get the
asymptotic value of the normalized probability for such processes
in the form
\begin{equation} \label{asymp_NC}
\begin{split}
 W_{{\rm asym}}  = & \sum\limits_{i = 1}^3 {\left| {U_{1i} } \right|^4 }  - \sum\limits_{i = 1}^3 {\left| {U_{1i} } \right|^4 \sin ^2 \delta _i } \ +  \\
 &\, + C_{\rm nc}\left( {1 - \sum\limits_{i = 1}^3 {\left| {U_{1i} } \right|^2 \sin ^2 \delta _i }  - \sum\limits_{i = 1}^3 {\left| {U_{1i} } \right|^4 }  + \sum\limits_{i = 1}^3 {\left| {U_{1i} } \right|^4 \sin ^2 \delta _i } } \right) ,
\end{split}
\end{equation}
where $\delta_i$ are the phases defined in Eq.~(\ref{asymp}) and
the coefficient $C_{\rm nc}$ is given by
\begin{equation}
C_{\rm nc}  = \int\limits_{\left| {\vec p} \right|_{\min } }^{\left|
{\vec p} \right|_{\max } } {\frac{{{\rm d}^3 W_{\rm P} }}{{{\rm
d}^3 p}}\, W_{\nu _\mu  e} \left| {\vec p} \right|^2 {\rm d} |
{\vec p} |} \, \left( \int\limits_{\left| {\vec p} \right|_{\min }
}^{\left| {\vec p} \right|_{\max } } {\frac{{{\rm d}^3 W_{\rm P}
}}{{{\rm d}^3 p}}\, W_{\nu _e  e} \left| {\vec p} \right|^2 {\rm
d} | {\vec p} |} \right)^{-1} ,
\end{equation}
$W_{\nu _\alpha  e}$ being the scattering probability of a
massless neutrino flavor state $\alpha$ at an electron, calculated
within the framework of the Standard Model. The term with $C_{\rm
nc}$ takes into account the contribution of the neutral current.
Numerical evaluation of $C_{\rm nc}$ for the neutrino production
in the $^{15}\rm O$ decay in our approximation gives $C_{\rm nc} =
0.210$. Using the explicit form of $W_{\nu _\alpha  e}$, one can
estimate  $C_{\rm nc}$ in the interval $420\, {\rm keV} \leq
\left| {\vec p} \right|_{\min } < \left| {\vec p} \right|_{\max }
\leq 14\, {\rm MeV}$ as  $0.177 < C_{\rm nc} < 0.321$.

\begin{figure}[h]
\begin{minipage}[h]{0.49\linewidth}
\begin{center}
\includegraphics[width=1\linewidth]{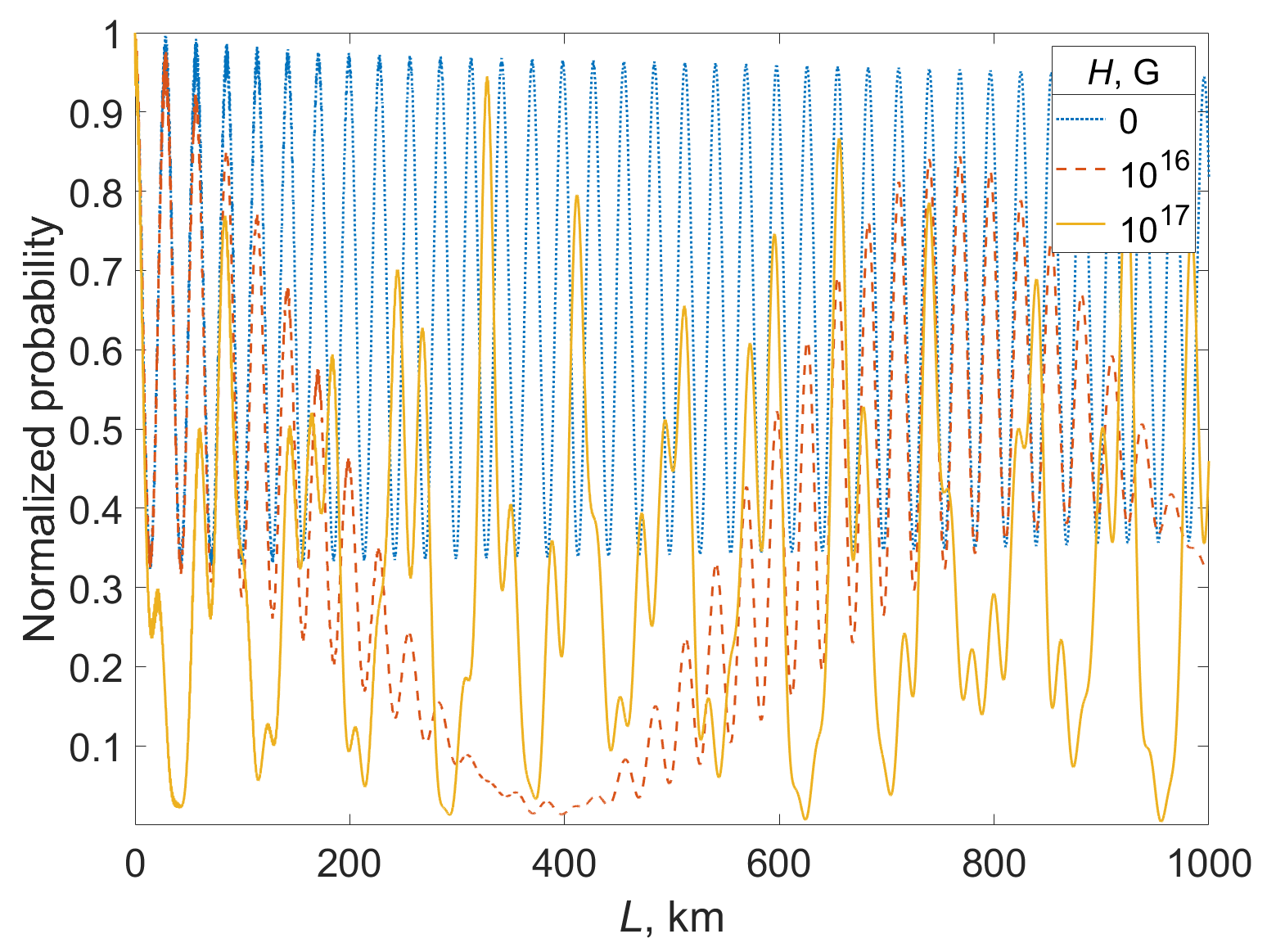}
\\ a) The first set of masses (\ref{set1}).
\end{center}
\end{minipage}
\hfill
\begin{minipage}[h]{0.49\linewidth}
\begin{center}
\includegraphics[width=1\linewidth]{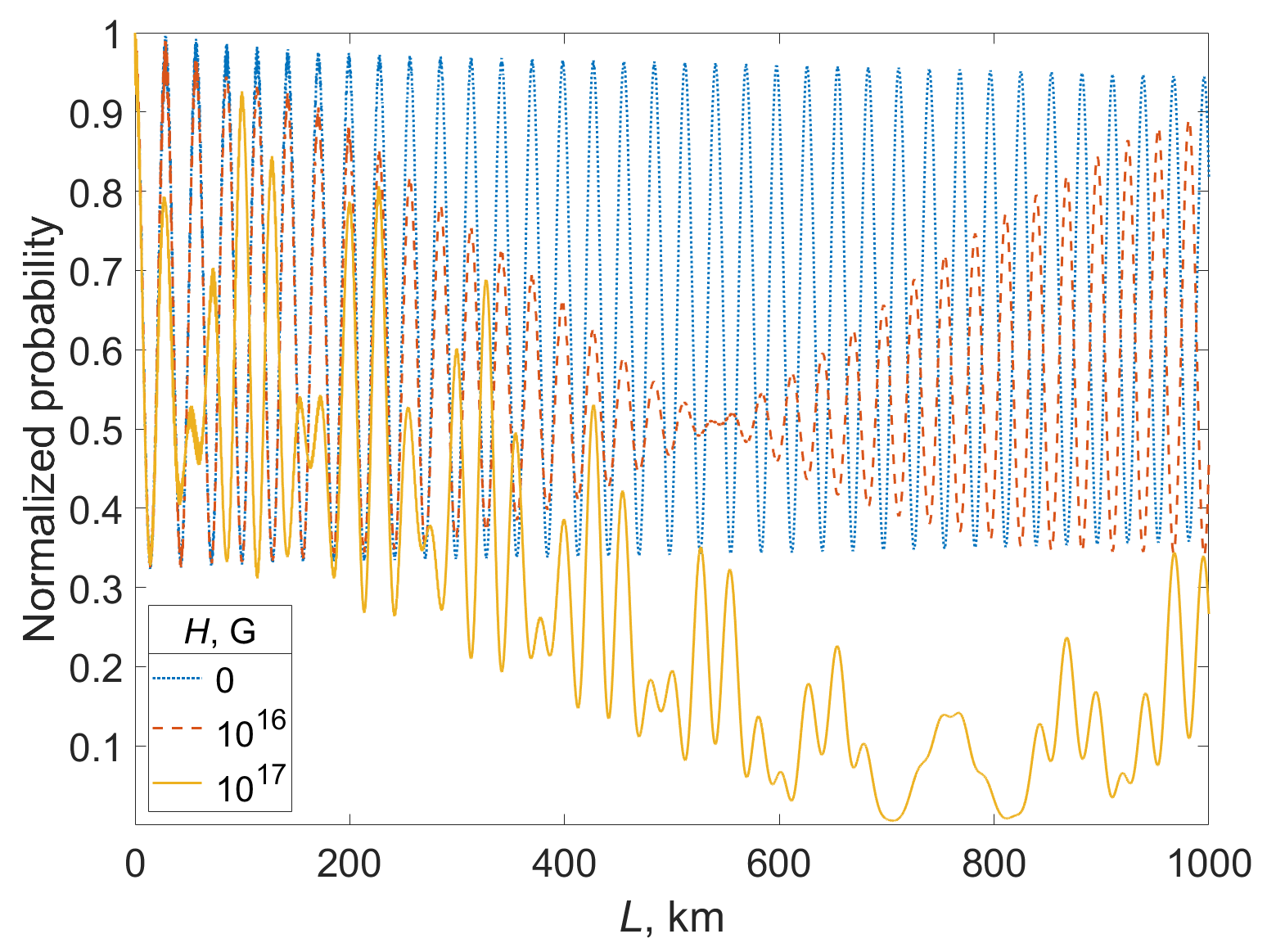}
\\ b) The second set of masses (\ref{set2}).
\end{center}
\end{minipage}
\caption{Averaged normalized probabilities (\ref{aver_prob_Be_NC})
of the  neutrino oscillation processes in constant homogeneous
magnetic field of various magnitude; neutrino production in the
electron capture by ${}^7{\rm Be}$ and detection by a Cherenkov
detector.} \label{osc_Be7_Cher}
\end{figure}

Finally, let us  consider  production reaction (\ref{Be7}) of
electron capture by ${^{7} {\rm Be}}$. Similar to the case of only
the charged-current interaction, averaging over the electron
momentum distribution in the source, we get the process
probability in the form
\begin{equation} \label{aver_prob_Be_NC}
\overline {\frac{{{\rm d}W}}{{{\rm d}\Omega }}}  = C\int\limits_\Delta ^\infty  { {\rm e}^ { - \frac{ \left| {\vec p} \right| -\Delta }{{kT}}} \sqrt {\left| {\vec p} \right| - \Delta } \, W_{\rm D} \left( {L,\overline H  } \right) \left| {\vec p} \right|^2 {\rm d} |{\vec p}| } .
\end{equation}
Assuming the plasma temperature to be $T = 1.5 \cdot 10^7$ K and
performing the numerical integration, we obtain the plots
presented in Fig.~\ref{osc_Be7_Cher}.

For this process with the solar neutrinos the coherence length is
also of the order of tens of thousand kilometers, and the
asymptotic value of the normalized probability is given by the
same expression (\ref{asymp_NC}), but the coefficient $C_{\rm nc}$
is calculated with the neutrino production probability in the
electron capture by ${^{7} {\rm Be}}$. The numerical integration
gives $C_{\rm nc} = 0.224$, which corresponds to the asymptotic
value  in vacuum $W_{{\rm asym}} = 0.652$. The ratio of the
neutrino flux measured by the Borexino collaboration to that
predicted for ${^{7} {\rm Be}}$ by the standard solar model is
$0.62 \pm 0.05$ \cite{Bellini:2011rx}. This is again in a good
agreement with our theoretical value and leaves room for a
contribution from the solar magnetic field, which is always
negative.

\section{Conclusion}
In the present paper we have shown that it is possible to give  a
consistent quantum field-theoretical description of neutrino
oscillations in a magnetic field in the Standard Model minimally
extended by the right neutrino singlets without use of the
neutrino flavor states. The description is performed in terms of
plane waves and is based on the Feynman diagram technique in the
coordinate representation supplemented with modified rules of
passing to the momentum representation. These rules reflect the
experimental setting and give rise to the distance-dependent
propagators of neutrino mass eigenstates. The distance-dependent
propagators in a magnetic field are explicitly calculated and
found to split into the sum of two terms corresponding to two
possible neutrino spin orientations and energies.

Processes of neutrino oscillation have been considered, where the
neutrinos  are  produced and detected through the charged- and
neutral-current weak interactions with nuclei and electrons  in
the absence of magnetic field, but the propagation of the
neutrinos takes place in a region of  magnetic field. Formulas for
the probabilities of the oscillation processes have been derived
and an agreement with the results obtained in the standard
quantum-mechanical description is shown.

An important new result derived within the
framework of the approach is formulas (\ref{asymp}) and
(\ref{asymp_NC}) for the asymptotic values of the normalized
probability of processes, where neutrinos are detected through the
interaction with nuclei or electrons. For a particular process,
this value is the observable ratio of the measurable neutrino flux
to that predicted by the standard solar model. These formulas can
be immediately compared with the experimental data. For the
neutrino detection through the interaction with electrons, also
limits on the neutral current contribution to the asymptotic value
are set.

The neutrino production in the solar core in
$^{15}\rm O$ decay and electron capture by ${^{7} {\rm Be}}$  and
detection by Ga-Ge and Cherenkov detectors has been studied.
Numerical computations of the normalized probabilities of these
processes have been performed and the results have been found to
be in a good agreement with the experimental data.

The advantages of the approach are the technical simplicity and
physical transparency. It makes use of only the standard tools of
perturbative quantum field theory and does not need wave packets
and the neutrino flavor states. In fact, we have shown that the
Standard  Model extended by the right neutrinos  is capable of
describing not only scattering processes, but also quantum
processes passing at finite space and time intervals, like
particle oscillations, in the framework of the standard Feynman
diagram technique and the modified perturbative formalism.

Finally, we have to note that we have not touched upon the problem
of neutrino oscillations in matter and  have chosen the examples
of the production and detection processes in the energy range,
where the influence of oscillations in matter is expected to be
weak. A quantum field-theoretical description of oscillations in
matter can be developed along the lines set forth in the present
paper, although the problem is rather complicated, because one has
to invert analytically a 12$\times$12 matrix to find Green's
function of neutrino mass eigenstates in matter.
Such a description is rigorous and very different from the
standard one. Hence the results may also differ. This is a matter
of a special investigation.

\section*{Acknowledgments}
The authors are grateful to E.~Boos, A.~Lobanov, A.~Pukhov
and Yu.~Tchuvilsky for interesting and useful
discussions. Special thanks are due to M.~Smolyakov for reading
the manuscript. Analytical calculations of the amplitudes have been
carried out with the help of the \mbox{CompHEP} and \mbox{Reduce} packages. The
work of V. Egorov was supported by the Foundation for the
Advancement of Theoretical Physics and Mathematics ``BASIS''.


\begin{thebibliography}{33}

\bibitem{Pontecorvo:1957cp}
  B.~Pontecorvo,
  Mesonium and anti-mesonium,
  Sov.\ Phys.\ JETP {\bf 6}, 429 (1957)
  [Zh.\ Eksp.\ Teor.\ Fiz.\ {\bf 33}, 549 (1957)].

\bibitem{Gribov:1968kq}
  V.~N.~Gribov and B.~Pontecorvo,
  Neutrino astronomy and lepton charge,
  Phys.\ Lett.\ B {\bf 28}, 493 (1969).

\bibitem{Giunti:2007ry}
  C.~Giunti and C.~W.~Kim,
  {\it Fundamentals of Neutrino Physics and Astrophysics}
  (Oxford University Press, Oxford, 2007).

\bibitem{Bilenky:2010zza}
  S.~Bilenky,
  Introduction to the physics of massive and mixed neutrinos,
  Lect.\ Notes Phys.\ {\bf 817}, 1 (2010).

\bibitem{Giunti:1993se}
  C.~Giunti, C.~W.~Kim, J.~A.~Lee and U.~W.~Lee,
  On the treatment of neutrino oscillations without resort to weak eigenstates,
  Phys.\ Rev.\ D {\bf 48}, 4310 (1993).

\bibitem{Grimus:1996av}
  W.~Grimus and P.~Stockinger,
  Real oscillations of virtual neutrinos,
  Phys.\ Rev.\ D {\bf 54}, 3414 (1996).

\bibitem{Beuthe:2001rc}
  M.~Beuthe,
  Oscillations of neutrinos and mesons in quantum field theory,
  Phys.\ Rept.\ {\bf 375}, 105 (2003).

\bibitem{Cohen:2008qb}
  A.~G.~Cohen, S.~L.~Glashow and Z.~Ligeti,
  Disentangling Neutrino Oscillations,
  Phys.\ Lett.\ B {\bf 678}, 191 (2009).

\bibitem{Kayser}
  B.~Kayser, On the quantum mechanics of neutrino oscillation,
  Phys.\ Rev.\ D {\bf 24}, 110 (1981).

\bibitem{Okun1982}
  I.~Yu.~Kobzarev, B.~V.~Martemyanov, L.~B.~Okun and M.~G.~Shchepkin,
  Sum rules for neutrino oscillations,
  Sov.\ J.\ Nucl.\ Phys.\ {\bf 35}, 708 (1982).

\bibitem{Grimus:2019hlq}
W.~Grimus,
J. Phys. G \textbf{47} (2020) no.8, 085004

\bibitem{Cisneros:1971}
    A.~Cisneros,
    Effect of neutrino magnetic moment on solar neutrino observations,
    Astrophys.\ Space Sci.\ \textbf{10}, 87 (1971).

\bibitem{Fujikawa:1980yx}
    K.~Fujikawa and R.~Shrock,
    The Magnetic Moment of a Massive Neutrino and Neutrino Spin Rotation,
    Phys.\ Rev.\ Lett.\ \textbf{45}, 963 (1980).

\bibitem{Schechter:1981hw}
    J.~Schechter and J.~W.~F.~Valle,
    Majorana Neutrinos and Magnetic Fields,
    Phys.\ Rev.\ D \textbf{24}, 1883 (1981) [erratum: Phys.\ Rev.\ D
    \textbf{25}, 283 (1982)].

\bibitem{Okun1986}
    M.~B.~Voloshin, M.~I.~Vysotsky and L.~B.~Okun,
    Neutrino electrodynamics and possible consequences for solar neutrinos,
    Sov.\ Phys.\ JETP {\bf 64}, 446 (1986).

\bibitem{Akhmedov}
    E.~K.~Akhmedov and J.~Pulido,
    Solar neutrino oscillations and bounds on neutrino magnetic moment and solar magnetic field,
    Phys.\ Lett.\ B {\bf 553}, 7 (2003).

\bibitem{Popov:2019nkr}
  A.~Popov and A.~Studenikin,
  Neutrino eigenstates and flavour, spin and spin-flavour oscillations in a constant magnetic field,
  Eur.\ Phys.\ J.\ C \textbf{79}, no.2, 144 (2019).

\bibitem{Chukhnova:2019oum}
  A.~V.~Chukhnova and A.~E.~Lobanov,
  Neutrino flavor oscillations and spin rotation in matter and electromagnetic field,
  Phys.\ Rev.\ D \textbf{101}, no.1, 013003 (2020).

\bibitem{Volobuev:2017izt}
  I.~P.~Volobuev,
  Quantum field-theoretical description of neutrino and neutral kaon oscillations,
  Int.\ J.\ Mod.\ Phys.\ A {\bf 33}, no.13, 1850075 (2018).

\bibitem{Egorov:2017qgk}
  V.~O.~Egorov and I.~P.~Volobuev,
  Neutrino oscillation processes in a quantum-field-theoretical approach,
  Phys.\ Rev.\ D {\bf 97}, no.9, 093002 (2018).

\bibitem{Egorov:2017vdp}
  V.~O.~Egorov and I.~P.~Volobuev,
  Neutrino oscillation processes with a change of lepton flavor in quantum field-theoretical approach,
  J.\ Exp.\ Theor.\ Phys.\ {\bf 128}, no.5, 713 (2019).

\bibitem{Volobuev:2019zan}
  I.~P.~Volobuev and V.~O.~Egorov,
  Quantum field theory description of processes passing at finite space and time intervals,
  Theor.\ Math.\ Phys.\ {\bf 199}, no.1, 562 (2019).

\bibitem{Egorov:2019vqv}
  V.~O.~Egorov and I.~P.~Volobuev,
  Coherence length of neutrino oscillations in a quantum field-theoretical approach,
  Phys.\ Rev.\ D \textbf{100}, no.3, 033004 (2019).

\bibitem{BOSH}
  N.~N.~Bogoliubov and D.~V.~Shirkov,
  {\it Introduction to the theory of quantized fields}, 3rd ed.
  (John Wiley \& Sons, New York, 1980).

\bibitem{Fujikawa:2020mei}
    K.~Fujikawa,
    Path integral of neutrino oscillations,
    arXiv:2009.08082 [hep-ph].

\bibitem{Ternov1965}
  I.~M.~Ternov, V.~G.~Bagrov and A.~M.~Khapaev,
  Electromagnetic radiation from a neutron in an external magnetic field,
  Sov.\ Phys.\ JETP \textbf{21}, no.3, 613 (1965).

\bibitem{BK}
  E.~Byckling and K.~Kajantie,
  {\it Particle Kinematics}
  (John Wiley \& Sons, London, 1973).

\bibitem{Bohr-Mottelson}
  A.~Bohr and B.~R.~Mottelson,
  {\it Nuclear Structure: Volume I: Single-Particle Motion}
  (World Scientific, Singapore, 1998).

\bibitem{Zyla:2020zbs}
  P.~A.~Zyla \textit{et al.} (Particle Data Group),
  Review of Particle Physics,
  PTEP \textbf{2020}, no.8, 083C01 (2020).

\bibitem{Aghanim:2018eyx}
  N.~Aghanim \textit{et al.} (Planck),
  Planck 2018 results. VI. Cosmological parameters,
  Astron.\ Astrophys.\ \textbf{641}, A6 (2020).

\bibitem{Okun1985}
  L.~B.~Okun,
  {\it Leptons and Quarks}
  (North Holland, 1985).

\bibitem{Borexino:2017fbd}
  M.~Agostini \textit{et al.} (Borexino),
  Limiting neutrino magnetic moments with Borexino Phase-II solar neutrino data,
  Phys.\ Rev.\ D \textbf{96}, no.9, 091103 (2017).

\bibitem{Beda:2012zz}
  A.~G.~Beda \textit{et al.},
  The results of search for the neutrino magnetic moment in GEMMA experiment,
  Adv.\ High Energy Phys.\ \textbf{2012}, 350150 (2012).

\bibitem{Bahcall:2000nu}
    J.~N.~Bahcall, M.~H.~Pinsonneault and S.~Basu,
    Solar models: Current epoch and time dependences, neutrinos, and helioseismological properties,
    Astrophys.\ J.\ \textbf{555}, 990 (2001).

\bibitem{Bellini:2011rx}
    G.~Bellini \textit{et al.},
    Precision measurement of the 7Be solar neutrino interaction rate in Borexino,
    Phys.\ Rev.\ Lett.\ \textbf{107}, 141302 (2011).

\end{thebibliography}
\end{document}